# High-throughput computational screening of small, eco-friendly, molecular crystals for sustainable piezoelectric materials

Shubham Vishnoi[1], Geetu Kumari[1], Robert Guest[1,2], Pierre-André Cazade[3*],

& Sarah Guerin[1, 2*]

[1]*Department of Chemical Sciences, Bernal Institute, University of Limerick, V94 T9PX, Ireland*

[2]*SSPC, The Science Foundation Ireland Research Centre for Pharmaceuticals, University of Limerick, V94 T9PX, Ireland*

[3]*Department of Physics, Bernal Institute, University of Limerick, V94 T9PX, Ireland*

**Abstract**

Organic molecular crystals are ideally placed to become next-generation piezoelectric materials due to their diverse chemistries that can be used to engineer tailor-made solid-state assemblies. Using crystal engineering principles, and techniques such as co-crystallisation, these materials can be engineered to have a wide range of electromechanical properties. For materials that have been structurally characterised by methods such as X-Ray Diffraction, computational chemistry is an effective tool to predict their electromechanical properties, allowing researchers to screen these molecular crystals and identify materials best suited to their chosen application. Here we present our database of small molecular crystals, and their Density Functional Theory (DFT) predicted electromechanical properties, CrystalDFT (https://actuatelab.ie/CrystalDFT). We highlight the broad range of electromechanical properties amongst this primary dataset, and in particular, the high number of crystals that have a naturally occurring longitudinal $d_{33}$ constant. This longitudinal electromechanical coupling is a prerequisite for several conventional sensing and energy harvesting applications, the presence of which is notably rare amongst the literature on biomolecular crystal piezoelectricity to date.

Piezoelectricity, a phenomenon where mechanical energy is converted into electrical energy and vice versa, is ubiquitous in our modern lives[1]. It finds applications in a wide range of devices, from diesel fuel injectors and electronic toothbrushes to medical ultrasound equipment. The dominant piezoelectric material in many applications is lead zirconate titanate (PZT)[2] and its derivatives, such as lead lanthanum zirconate titanate (PLZT)[3], valued for their excellent piezoelectric coefficients and electromechanical coupling factors. However, the use of lead in PZT poses environmental and health concerns[4], leading the European Union to designate lead zirconate titanate as a hazardous substance. Consequently, there is a growing imperative to explore alternative materials that can offer comparable or superior performance without the environmental drawbacks associated with lead-based compounds or other ceramics[5-7]. This pursuit reflects the ongoing commitment to advancing sustainable and eco-friendly technologies in the realm of piezoelectric applications.

Organic biomolecular crystals are an emerging class of eco-friendly piezoelectric materials[8] with tuneable chemical properties that contribute to their potential applications. Additionally, they are biocompatible and biodegradable, making them suitable for lead-free biomedical[9] applications and smart devices while ensuring environmentally friendly production and disposal (**Figure 1**)[10]. The vast majority of biological materials naturally lack a centre of symmetry in their crystal structure, endowing them with piezoelectric properties[11,12]. In these non-centrosymmetric crystals, a surface charge is generated under an applied force due to ionic displacement. This phenomenon arises from the non-centrosymmetric structures of these biomolecules when they self-assemble and/or crystallise in the solid state, allowing them to exhibit piezoelectric properties like traditional inorganic materials. The unique molecular arrangements in organic crystals, driven by hydrogen bonding, van der Waals forces, and π-π interactions, allow for the design of tailored crystal structures that can exhibit piezoelectric properties[13].

In the past seventy years, piezoelectricity has been confirmed in a variety of biological materials including wood[14] and bone[15], and in fibrous proteins like collagen[16], chitin[17] and elastin[18], which present as highly ordered crystalline molecules in mammalian tissue. Small biomolecules, such as certain amino acids and dipeptides with non-centrosymmetric structures, exhibit significant piezoelectric properties due to their ability to generate net polarization (**Figure 1** and **Table S1A)**.

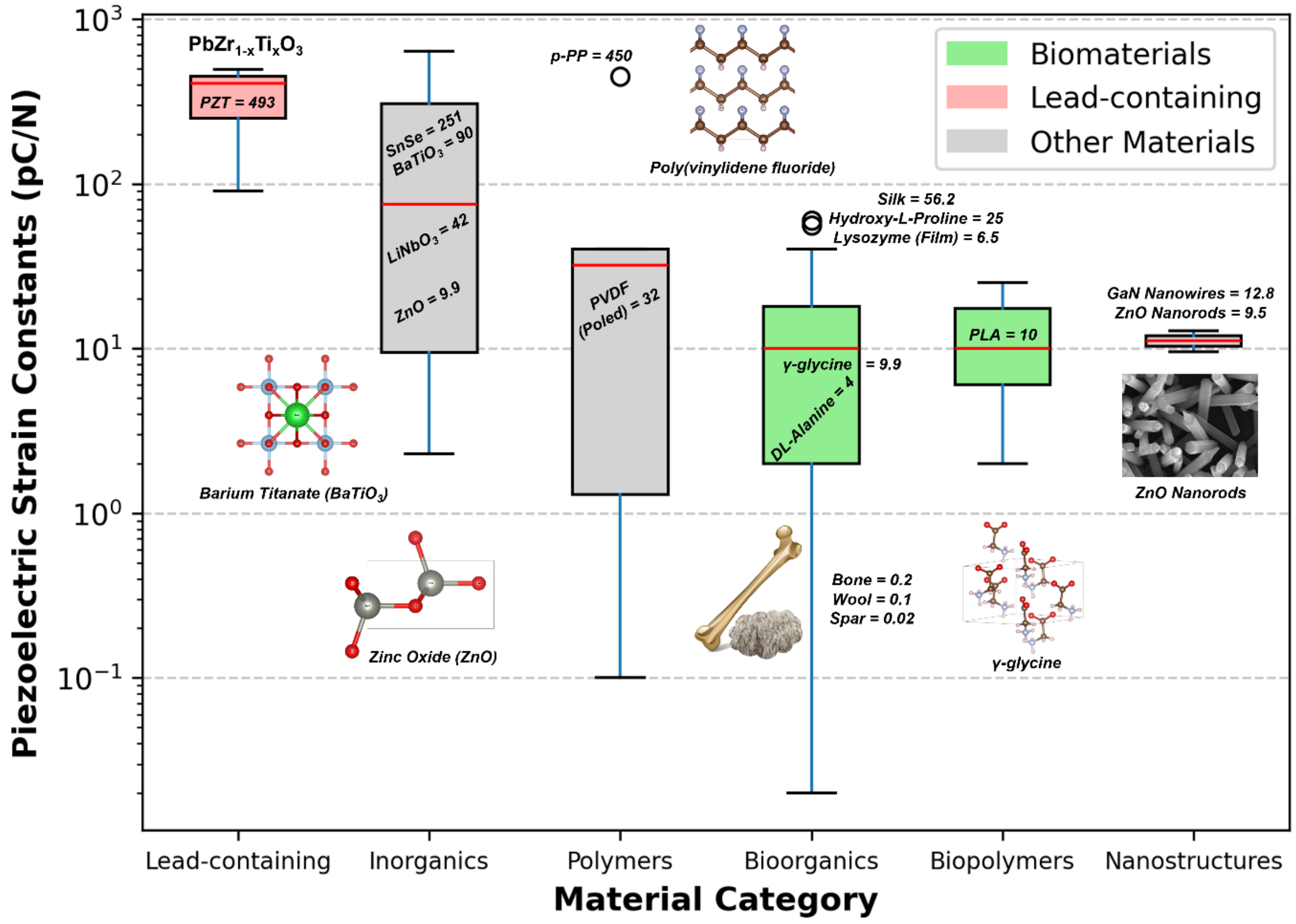

**Figure 1.** Comparison of piezoelectric strain coefficients in leading and emerging materials: a comparison is made between the piezoelectric strain coefficients of various bioorganic and organic materials *vs*. traditional inorganic materials and polymers. While bioorganic substances generally exhibit lower strain coefficients compared to the materials commonly utilised in electronic devices, their significantly lower dielectric constants result in exceptionally high voltage constants. This makes them particularly promising for applications such as energy harvesting and sensing. The error bars shown represent the range of values observed across the sampled materials, with detailed data sources provided in Supplementary Information **Table S1A**. The read line in the plot shows the median Piezoelectric Strain Coefficient (pC/N) for each category (**Table S1B**), highlighting differences in strain values and indicating their currently observed potential for different energy harvesting and sensing applications.

The introduction of Density Functional Perturbation Theory (DFPT) by Baroni *et al.* in 1987 enabled efficient computation of energy derivatives with respect to atomic displacements, strain and electric fields[19]. Since then, DFPT has been instrumental in the theoretical investigation of numerous piezoelectric materials[20-23]. First-principles calculations based on DFT have been widely utilised to predict material properties such as dielectric constants, piezoelectric constants, elastic constants, phonon dispersion relationships *etc*[24]. The utilisation of high-throughput DFT calculations for material screening and fundamental research offers a promising and innovative avenue in materials science[23,25]. DFT has revolutionised computational materials science by providing an efficient framework for predicting the properties of complex systems. High-throughput DFT extends this capability, enabling the rapid screening of vast chemical spaces and the discovery of new materials. High-throughput DFT involves analysing hundreds to tens of thousands of compounds, necessitating novel calculation and data management approaches. Applications of High-throughput DFT have led to the development of several comprehensive databases, including AFLOW[26], Materials Project[27], Open Quantum Materials Database (OQMD)[28]and NIST-JARVIS[29]. These databases catalogue computed geometries and various physicochemical properties, facilitating the exploration of new materials. Specific to piezoelectric[1,30,31] and dielectric[20,32] materials, databases developed have laid the groundwork for the resource provided in this work. Our work expands on these efforts by developing a systematic database for small, eco-friendly, molecular crystals for sustainable piezoelectric materials, significantly enhancing the scope of previous research[1,30,31].

For this resource, we have applied high-throughput screening to a curated database of organic crystals, to predict their anisotropic electromechanical behaviour. Through this approach, we have identified several organic crystals that exhibit promising piezoelectric properties, some of them exceeding the performance of well-known inorganic piezoelectric materials. These findings not only demonstrate the potential of bio/organic crystals for piezoelectric applications but also highlight the efficiency of high-throughput methods in accelerating material discovery and optimization. This resource provides the essential data infrastructure for scalable data generation and analysis in large-scale projects. It explores challenges in accurately computing properties across diverse chemical spaces using a single exchange-correlation functional, emphasising how variations in atomic arrangements contribute to errors in the generalized gradient approximation. To advance understanding of the structural features that dictate electromechanical outputs, we report consistently calculated piezoelectric tensors of these crystalline, non-centrosymmetric organic materials using our high-throughput DFT calculation workflow. Gaining insights into the electronic structure of these organic crystals not only supports future experimental work but also enhances the understanding of their piezoelectric behaviour.

## Results

The screening of organic piezoelectric materials was successfully conducted through our comprehensive computational high-throughput methods, utilising quantum mechanical modelling (QMM) *via* DFT calculations

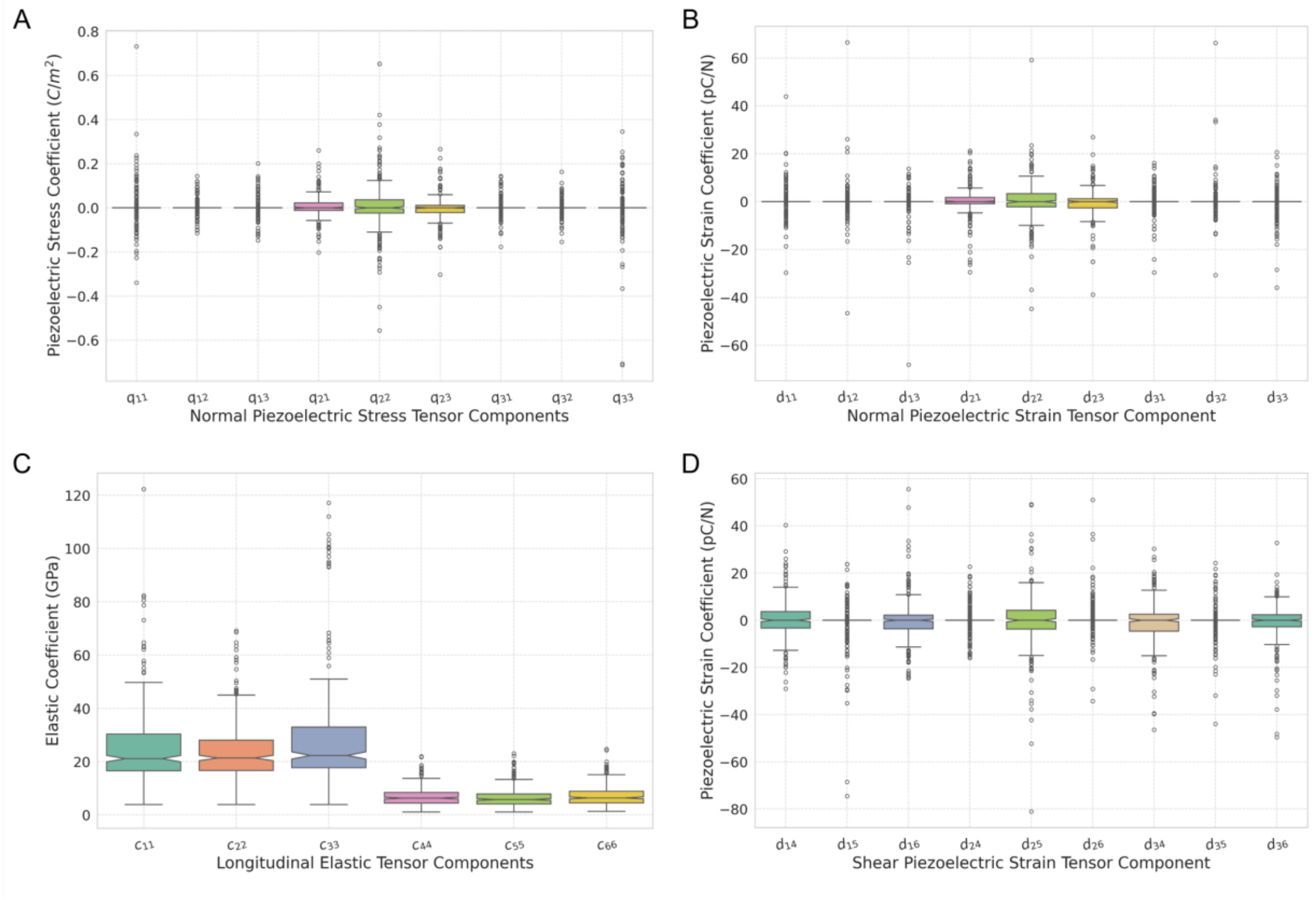

**Figure 2.** Predicted mechanical and piezoelectric properties of all crystals, excluding those flagged for high dielectric constants ($\varepsilon > 5$), negative material stiffness, high condition numbers (CN > 100) or negative eigenvalues.

**Figure 2** shows the full spectrum of predicted electromechanical properties for stable structures in the CrystalDFT Database. **Figure 2A** shows the predicted piezoelectric charge tensor components, a key Figure of Merit for ferroelectric applications, in C/m$^2$. The majority of values are less than ± 0.1 C/m$^2$, which is consistent with our previous small-scale investigations. It is promising however that the highest predicted values are longitudinal, making them easily exploitable for device applications. Longitudinal piezoelectricity indicates the presence of a dipole along an orthogonal direction, meaning that these properties could also be enhanced via poling. The highest predicted values approach 0.8 C/m$^2$, a record previously held amongst biomolecular crystals by gamma glycine. These values still fall short of the charge coefficients of ceramics and other classes of materials, the majority of which exceed 1 C/m$^2$.

**Figure 2B and D** shows the piezoelectric strain constants, which represent the ratio of the piezoelectric charge to the elastic stiffness (individual elastic stiffness tensor components are shown in **Figure 2C**). The data presented in **Figure 2** is for crystals with cautiously low coordination numbers, and as a result the highest predicted value approaches 80 pC/N. However, there are crystals in this database that pass all other flags with slightly higher co-ordination numbers that have triple digit responses, rivalling ceramics such as Barium Titanate.

It is challenging for organic crystals to demonstrate large piezoelectric strain constants in comparison to the highest performing commercially available piezoelectric materials such as PZT. However, a large number of materials that we have identified in this work have $d_{33}$ constants significantly higher than those of many commonly used piezoelectric materials like – Quartz, ZnO, AlN etc. **Figure 2C** provides an overview of the primary longitudinal and shear elastic stiffness constants, providing statistical evidence of the low shear stiffness in this material class, which often results in high shear piezoelectricity[33-35]. It is interesting that the majority of small molecular crystals in this work have longitudinal stiffness between 20 and 40 GPa, which is a contrast to widely-reported flexible molecular crystals[36,37].

The top organic piezoelectric material candidates represent a subset of the most promising crystals derived from the computational screening process (**Figure 3**). A total of 572 crystals were evaluated, leading to the identification of 22 candidates exhibiting double-digit piezoelectric coefficients measured in picocoulombs per Newton (pC/N). The ranked crystals are exclusively from materials that met stringent stability and performance criteria (**Supplementary Note S1**). The results of the screening process are summarised in **Figure 3**. **Figure 3A** illustrates the direct piezoelectric effect, where mechanical stress generates an electric charge, alongside the converse piezoelectric effect, which demonstrates how an applied electric field induces strain. The crystals screened in this work will demonstrate both the direct and converse piezoelectric effect.

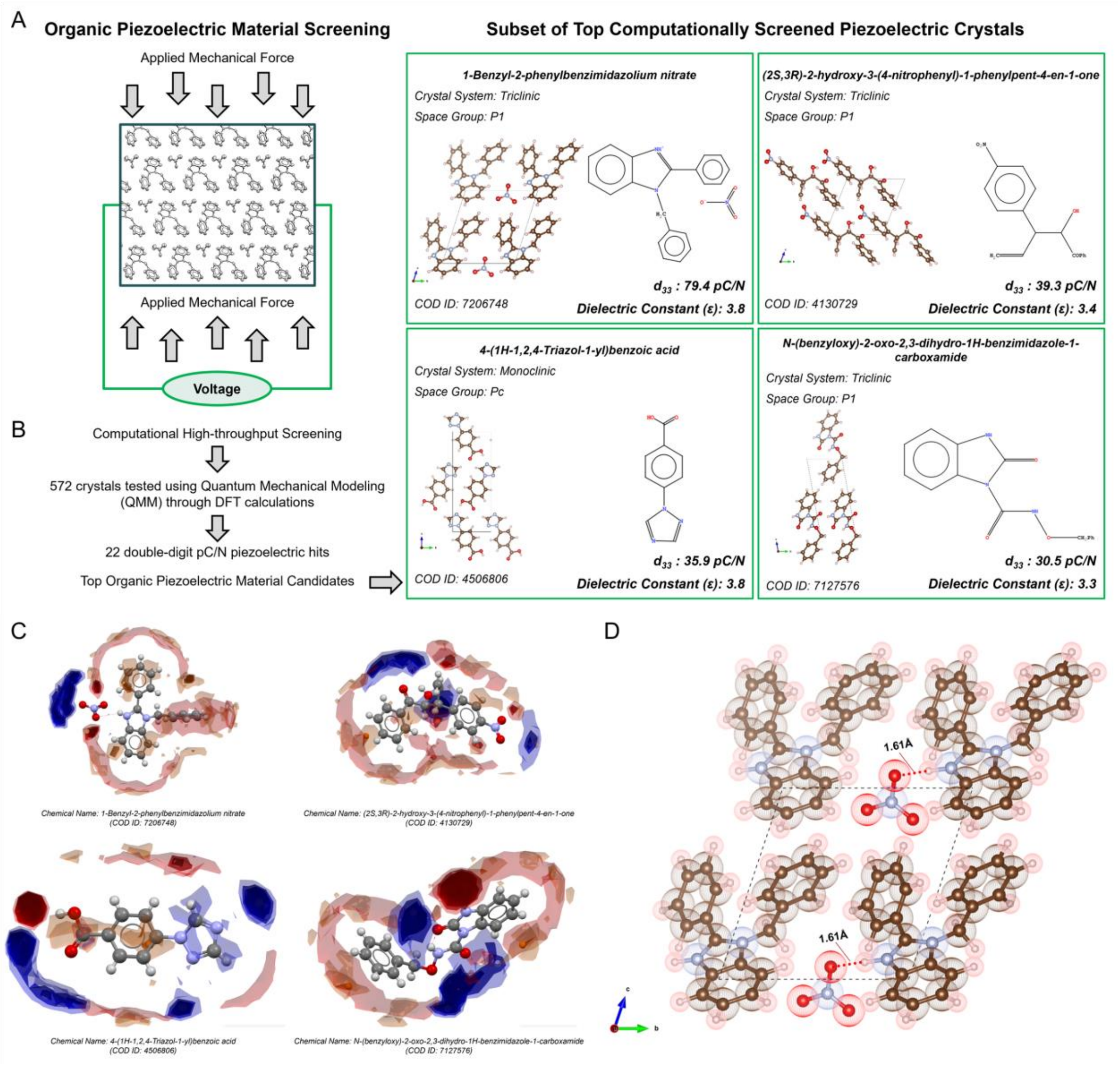

**Figure 3.** Figure provides a comprehensive summary of the organic piezoelectric material screening process using computational high-throughput methods. A total of 572 crystals were evaluated through density functional theory (DFT) calculations, resulting in the identification of 22 candidates with double-digit piezoelectric coefficients (pC/N), showcasing the most promising organic materials for potential development. (A) Illustrates a schematic representation of the direct piezoelectric effect, where mechanical stress generates an electric charge, and the converse piezoelectric effect, demonstrating strain induced by an applied electric field. (B) Summarises the results of the computational high-throughput screening, highlighting the top organic piezoelectric material candidates. (C) Features interaction maps around the molecular conformations of these candidates, with contour levels of 2, 4, and 6 shown with increasing opacity. Finally, (D) Displays the single-crystal structure of 1-Benzyl-2-phenylbenzimidazolium nitrate, notable for its high predicted longitudinal piezoelectric coefficient ($d_{33}$).

The top organic crystals were ranked based on their predicted longitudinal piezoelectric strain coefficient ($d_{33}$), the most widely exploited piezoelectric response for technological applications. In the initial assessment (**Figure 4A**), all evaluated crystals were included, highlighting a wide range of candidates with varying properties. However, to refine the selection, crystals with a dielectric constant greater than 5 ($\varepsilon > 5$) or exhibiting negative material stiffness were excluded, as we use these exclusion criteria to identify structures with structural abnormalities e.g. missing hydrogen atoms (**Table 1**). The refined list (**Figure 4B**) shows the most viable candidates, emphasising those with optimal piezoelectric properties and stable material characteristics. **Figure 3B** summarises the computational screening results, showcasing the most promising organic piezoelectric material candidates identified during the evaluation (**Table 2**), with repeat testing and multiple replicates per system conducted to ensure adequate sampling and reproducibility (**Figure S1**). Additionally, **Figure 3C** features interaction maps surrounding the molecular conformations of these top candidates, with contour levels depicted using increasing opacity to indicate varying strengths of molecular interactions. This visualisation provides insights into the supramolecular packing motifs that endow strong piezoelectric properties. Finally, **Figure 3D** displays the single-crystal structure of 1-Benzyl-2-phenylbenzimidazolium nitrate, with a high predicted longitudinal piezoelectric coefficient ($d_{33}$) of 79.4 pC/N. **Figure 4. (A)** illustrates the $d_{33}$ coefficients of all the organic crystals in our database including flagged crystals. On the other hands, **(B),** represents all those structures which are not flagged which means that they match all the flagging criteria for creation of our database. This chart illustrates the top twenty crystals, with their $d_{33}$ values roughly in the range of approximately 11-80 pC/N, which exceed the piezoelectric response of many conventional piezoelectric materials such as Zinc Oxide[38], Aluminium Nitrate[39], and the polymer PVDF[40,41]. Out of these twenty crystals (**Table 1**), the maximum longitudinal piezo response value that has been observed is 79.4 pC/N (COD ID is – 7206748) and the minimum of belong to triclinic space group ($P_1$), which is the lowest symmetry space group, which thus explains that piezoelectricity is space-group dependent property.

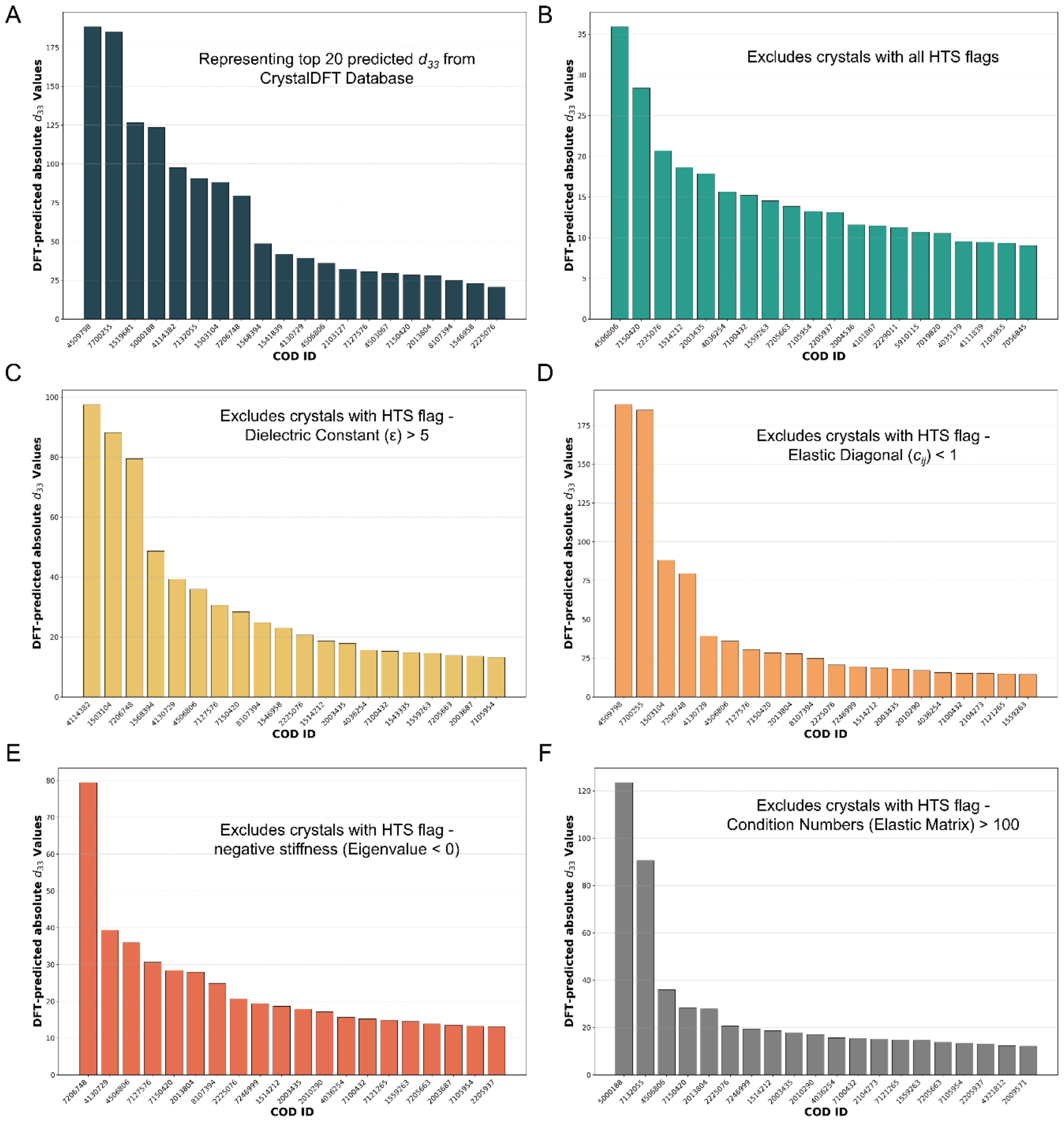

**Figure 4.** Top organic crystals ranked by DFT-predicted longitudinal piezoelectric strain coefficient ($d_{33}$). Bar chart showing the top 20 materials ranked by the absolute values of their predicted $d_{33}$ coefficients, indicating their potential for electromechanical applications. Only materials meeting stability and performance criteria are included in this analysis. Specifically, crystals with high dielectric constants, negative stiffness, or high condition numbers have been excluded to ensure accuracy and reliability of the data. The chart highlights the relative magnitude of $|d_{33}|$ values across selected materials, with each bar representing an individual crystal identified by its COD ID. These criteria allow focus on materials most suitable for practical applications while filtering out those that may be unstable or exhibit high susceptibility to dielectric loss. **(A)** Includes all crystals, while **(B)** Excludes crystals flagged for all four high-throughput flags (higher dielectric constant (ε > 5) or negative material stiffness (elastic tensor diagonal, *Cij* < 1 or Eigenvalue < 0), **(C)** Excludes crystals only flagged due to their higher dielectric constant **(D)** Excludes crystals only flagged due to their poor predicted elasticity **(E)** Excludes crystals only flagged due to their negative stiffness **(F)** Excludes crystals only flagged due to their high susceptibility towards elastic tensor components. All flag criteria are detailed in **Supplementary Note S1**.

**Table 1** shows the top-performing crystals with their COD ID, exhibiting high longitudinal piezoelectric constants ($d_{33}$) and low dielectric constants (unitless quantity). There are some negative $d_{33}$ values, those negative values of $d_{33}$ are because of the polarisation and hence by changing the polarity we can change the sign of the coefficients. All these crystals possess dielectric values below 5. The low dielectric constants of organic crystals have allowed them to excel as piezoelectric materials in the domains of energy harvesting. Lower dielectric values mean a lower device capacitance, reducing charge retention under an applied electric field. In a piezoelectric device, a lower capacitance allows for more efficient transfer of electrical energy when the material is mechanically stressed. This improved efficiency arises because less energy is "lost" in the form of stored charge, and allowing the material to generate a stronger piezoelectric response.

**Table 1.** Top organic crystals ranked by predicted longitudinal piezoelectric strain coefficient ($d_{33}$) along with their predicted dielectric constant ($\varepsilon$), excluding crystals flagged for higher dielectric constant ($\varepsilon > 5$) or negative material stiffness.

| Serial No. | COD ID | $d_{33}$ (pC/N) | Dielectric constant $\varepsilon$ |
|---|---|---|---|
| **1** | 7206748 | 79.4 | 3.8 |
| **2** | 4130729 | 39.3 | 3.4 |
| **3** | 4506806 | -35.9 | 3.8 |
| **4** | 7127576 | 30.5 | 3.3 |
| **5** | 7150420 | -28.3 | 2.9 |
| **6** | 8107394 | 24.8 | 4.2 |
| **7** | 2225076 | 20.6 | 3.4 |
| **8** | 1514212 | 18.6 | 2.9 |
| **9** | 2003435 | -17.8 | 3.3 |
| **10** | 4036254 | -15.6 | 2.9 |
| **11** | 7100432 | 15.2 | 3.1 |
| **12** | 1559263 | -14.5 | 3 |
| **13** | 7205663 | -13.8 | 3.9 |
| **14** | 7105954 | -13.2 | 4.8 |
| **15** | 2205937 | -13.1 | 3.7 |
| **16** | 7228079 | -12.7 | 3.6 |
| **17** | 4114383 | -11.9 | 3.1 |
| **18** | 2004536 | 11.5 | 3.3 |
| **19** | 4101867 | -11.4 | 3.3 |
| **20** | 2229011 | 11.2 | 4.3 |
| **21** | 5910115 | 10.7 | 2.8 |
| **22** | 7019820 | -10.5 | 3.5 |

The detailed information of the top five crystals is given in **Table 2.** that contains their COD ID, Space Group, Crystal System, Compound Name, Molecular Formula, Lattice Constants (in Å), Cell Volume $(Å)^3$, Cell Angle (in °), $d_{33}$ (pC/N).

**Table 2.** Details of top 5 crystals with predicted longitudinal piezoelectric strain coefficient values.

| COD ID | Space Group | Crystal System | Compound Name | Molecular Formula | Lattice Constants | | | | | | Cell Volume ($Å^3$) | *d33* (pC/N) |
|---|---|---|---|---|---|---|---|---|---|---|---|---|
| | | | | | Cell Length (Å) | | | Cell Angle (°) | | | | |
| | | | | | *a* | *b* | *c* | *α* | *β* | *γ* | | |
| 7206748 | P1 | Triclinic | *1-Benzyl-2-phenylbenzimidazolium nitrate* | $C_{20}H_{17}N_3O_3$ | 5.71 | 8.89 | 9.5 | 71.73 | 79.38 | 84.2 | 449.3 | 79.4 |
| 4130729 | P1 | Triclinic | *(2S,3R)-2-hydroxy-3-(4-nitrophenyl)-1-phenylpent-4-en-1-one* | $C_{17}H_{15}NO_4$ | 6.47 | 7.73 | 8 | 68.43 | 85.64 | 87.9 | 370.7 | 39.3 |
| 4506806 | Pc | Monoclinic | *4-(1H-1,2,4-Triazol-1-yl)benzoic acid* | $C_9H_7N_3O_2$ | 3.96 | 6.29 | 16.52 | 90 | 94.46 | 90 | 410.4 | -35.9 |
| 7127576 | P1 | Triclinic | *N-(benzyloxy)-2-oxo-2,3-dihydro-1H-benzimidazole-1-carboxamide* | $C_{15}H_{13}N_3O_3$ | 5.1 | 5.36 | 12.34 | 99.97 | 96.11 | 90.9 | 330.34 | 30.5 |
| 7150420 | P1 | Triclinic | *Boc-betaPro-OH* | $C_{11}H_{19}NO_4$ | 5.99 | 6.65 | 8.66 | 70.75 | 77.42 | 86.98 | 317.81 | -28.3 |

The analysis of the CrystalDFT dataset highlights a wide range of molecular characteristics, focusing particularly on functional groups that could enhance the piezoelectric properties of the materials (**Figure 5A, 6B and C**). Among the 9 functional groups and backbone substructures examined, aldehyde and hydroxyl followed by amino groups exhibited the highest presence, with counts of 107, 106 and 98, respectively, indicating a robust foundation for various chemical interactions. The hydroxyl group, often associated with hydrogen bonding and enhanced polarity, plays a crucial role in influencing the stability and performance of the crystals. Similarly, an analysis of the backbone substructure reveals that alkyl groups, including alkanes and alkenes, as well as aromatic rings, with their substantial representation in the dataset, suggest a strong correlation with piezoelectric characteristics, likely due to their stable molecular structure and their favourable hydrophobic characteristics within the crystals (**Figure 5B**). The piezoelectric responses associated with other functional groups are particularly significant. For instance, both the amino and aldehyde groups exhibit substantial stable subset counts of 151 and piezoelectric counts of 107 and 98, highlighting their potential for further exploration. Functional groups with high presence in the unflagged organic crystals subset, such as alkane, amino, aldehyde, hydroxyl, ketone, carboxylic acid groups and aromatic rings, are strongly associated with stable piezoelectric properties (**Figure 6C**). Their significant representation, after excluding crystals with high dielectric constants, negative stiffness, or high condition numbers, underscores their reliability in enhancing electromechanical performance.

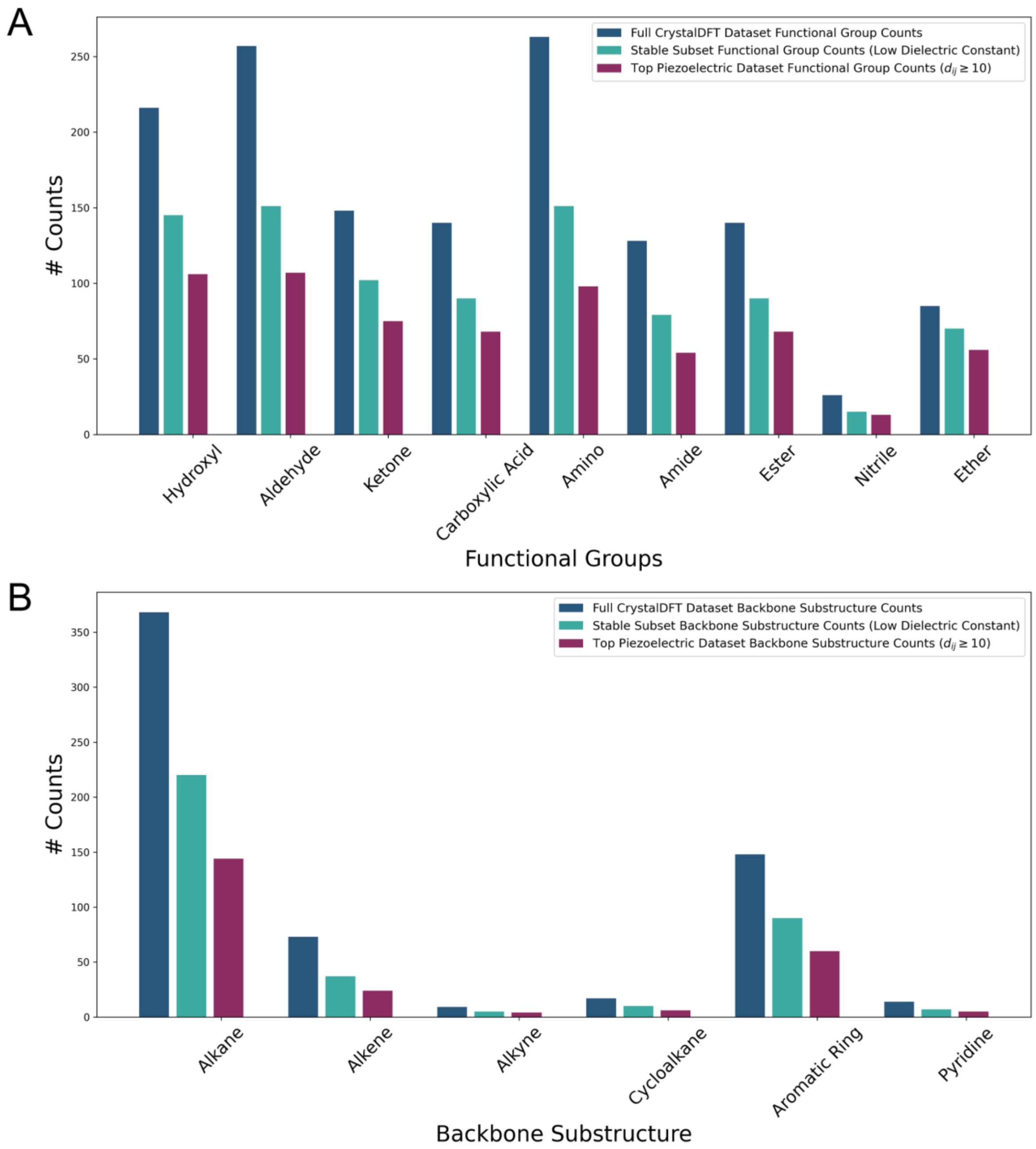

**Figure 5.** Analysis on the CrystalDFT dataset showcasing the diversity of molecular characteristics, with a particular emphasis on functional groups and backbone substructures. **(A)** Functional groups such as aldehyde, hydroxyl and amino groups may contribute to the piezoelectric properties of stable organic materials. **(B)** While considering alkanes as the basic backbone, aromatic ring structures might function as potential substructures for designing organic piezoelectric materials, offering valuable insights into the structure-property relationships within these materials. Further details are provided in **Supplementary Table S2**.

Furthermore, only a significant proportion of the crystals possess a naturally occurring longitudinal $d_{33}$ constant, a critical feature for effective electromechanical applications, and hence this functional group analysis should not be generalised to other space group symmetries with zero $d_{33}$ component. This finding underscores the broad spectrum of electromechanical properties present in the primary dataset and emphasises how particular molecular features can influence piezoelectric performance. Through the lens of HTS flags, the subset of crystals that met these rigorous criteria showcases materials with promising electromechanical capabilities. Examining these functional groups reveals not only their distinct chemical properties but also offers insights into how molecular structure could be correlated with piezoelectric behaviour, paving the way for future investigations aimed at refining crystal design for advanced piezoelectric applications.

**Discussion: Outlook for the CrystalDFT Project**

Molecular crystals with inherent piezoelectric properties offer a promising alternative to currently used lead-based piezoelectric materials. Lead-based materials, such as lead zirconate titanate (PZT), have been widely utilised for their excellent piezoelectric performance but raise environmental and health concerns due to the toxicity of lead. This approach allows us to shift through a vast array of potential candidates efficiently, identifying materials that not only meet but surpass the requirements for various applications such as sensing and actuation. Advancing materials science through computational method development will open new avenues for the development of cutting-edge materials with enhanced performance, with impact on a diverse set of industries along with accelerated technological advancement.

All of the top four low-CN candidates shown in **Figure 3** contain a phenyl and/or benzene derivative component, which is consistent with the large number of studies on piezoelectricity on di-phenylalanine and similar molecules that facilitate pi-pi stacking interactions[42-44]. The next highest ranked crystal on the list is *Boc-betaPro-OH*, which is chemically similar to the hydroxy-L-proline based amino acids and peptides that we have observed high piezoelectricity and voltage output from in previous work[45,46].

In this study, we focused specifically on small organic crystals, which are of particular interest due to their diverse molecular architectures and tuneable properties. These small organic crystals often display the same lack of centrosymmetry, thereby exhibiting piezoelectricity. By concentrating on smaller molecular systems (**Figure 6**), we sought to provide a resource to uncover new organic piezoelectric materials with practical applications, taking advantage of their simpler structures and ease of synthesis compared to larger, more complex biological systems. DFT allows for the systematic analysis of these materials' electronic structures and mechanical responses, enabling rapid identification of promising candidates (**Figure 2**). Screening small organic crystals, in particular, is advantageous, as their structural diversity often leads to a wide range of electromechnical behaviours, enhancing the chances of discovering materials with high piezoelectric coefficients and desirable properties for future applications.

This database is significant as it provides a resource to the community to identify and screen organic crystals for any electromechnical application, contributing to the development of sustainable materials for electronic applications. By employing high-throughput DFT methods, we efficiently explore a diverse set of organic compounds, accelerating the discovery process. These high values suggest that organic crystals could be more effective at producing piezoelectricity than many key inorganic crystals. By continuing to explore databases of organic materials, new and better piezoelectric materials for various applications, such as sensors, actuators, and energy-harvesting devices could be discovered. We look forward to using this workflow to study larger and more complex peptide and protein crystals, with novel applications in tissue regeneration, drug delivery, and antimicrobial coatings. Moving forward, by training machine learning models on databases like CrystalDFT and the Materials Project, we can make better use of available information, further enhancing the accuracy and efficiency of piezoelectric material discovery.

Our analysis demonstrates the efficacy of the chosen computational model in predicting molecular properties. The interpretability of the model provides valuable insights for chemists and researchers, contributing to a deeper understanding of the relationships between molecular features and the target property.

## Online Methods

*Database Development:* We have curated a dataset of ~572 non-centrosymmetric organic structures from the *Crystallographic Open Database (COD)*[47,48]. We used screening criteria to curate the target organic crystals that have the possibility of exhibiting piezoelectric behaviour (**Figure 6A** and **Supplementary Notes S1/2**). Only structures with space groups 1, 3–9, 16–46, 75–82, 89–122, 143–146, 149–161, 168–174, 177–190, 195–199, 207–220 were considered since these space groups lack inversion symmetry (**Table S3**). We set an initial limit of each organic crystal structure consisting of 50 atoms per unit cell in order to maximize the number of calculable structures within the available computational time. We have focused on investigating the piezoelectric properties of these organic crystals through high-throughput DFT calculations. To benchmark the piezoelectric properties of these simulated organic crystals and to validate overall high-throughput workflow, we also performed DFT calculations on a set of well-known crystal structures as listed in **Table S4**. Our reference piezoelectric materials dataset includes zinc oxide, aluminium nitride, α-quartz, barium titanate and additional materials for comprehensive comparison.

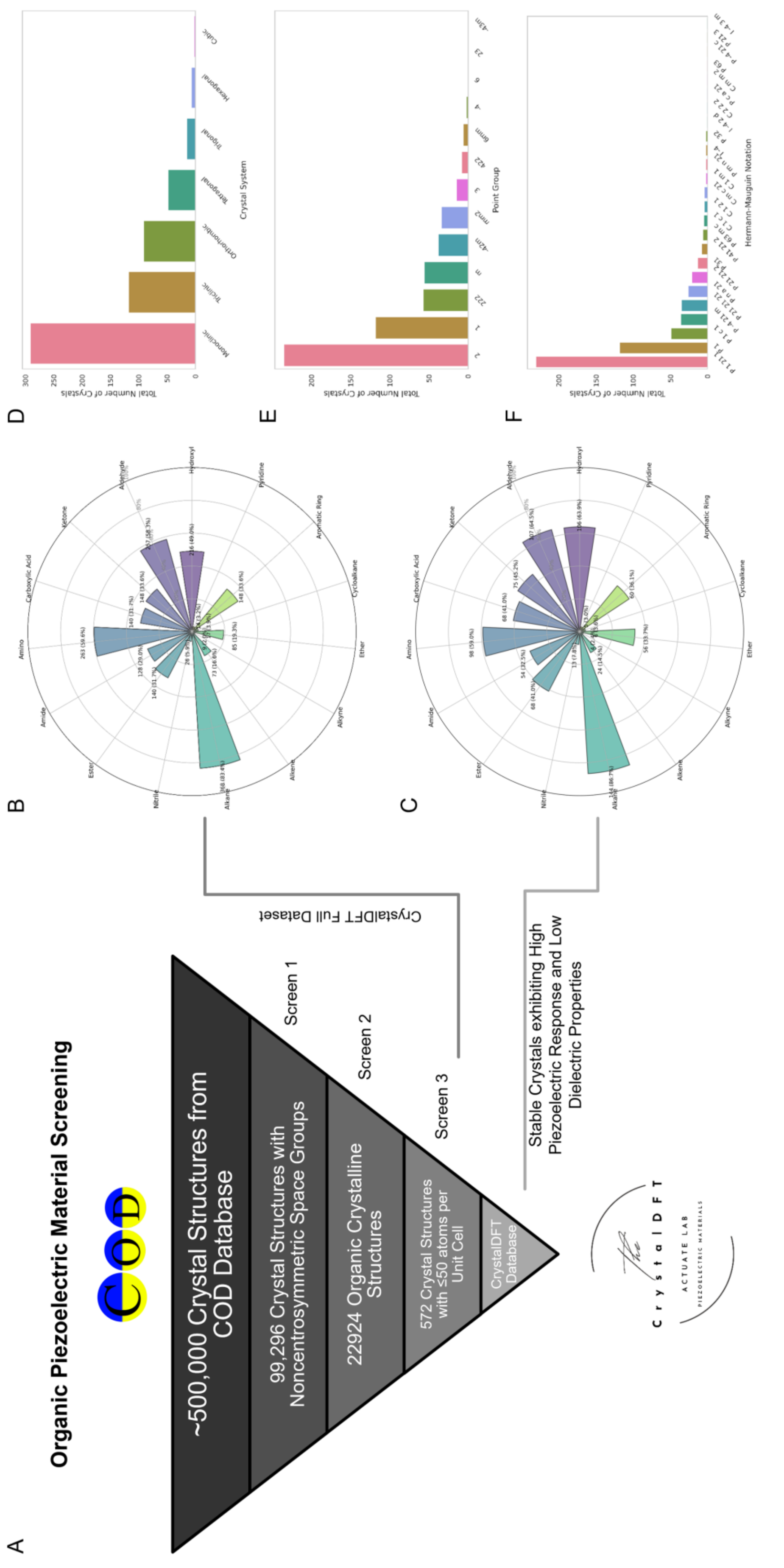
A
Organic Piezoelectric Material Screening
COD
~500,000 Crystal Structures from COD Database
99,296 Crystal Structures with Noncentrosymmetric Space Groups
22924 Organic Crystalline Structures
572 Crystal Structures with ≤50 atoms per Unit Cell
CrystalDFT Database
Screen 1
Screen 2
Screen 3
CrystalDFT Full Dataset
Stable Crystals exhibiting High Piezoelectric Response and Low Dielectric Properties
The CrystalDFT
ACTUATE LAB
PIEZOELECTRIC MATERIALS
B
C
Hydroxyl
Pyridine
Aromatic Ring
Cycloalkane
Ether
Alkyne
Alkene
Alkane
Nitrile
Ester
Amide
Amino
Carboxylic Acid
Ketone
Aldehyde
D
E
F
Total Number of Crystals
Crystal System
Monoclinic
Triclinic
Orthorhombic
Tetragonal
Trigonal
Hexagonal
Cubic
Point Group
Hermann-Mauguin Notation

**Figure 6.** Overview of the computational screening process for organic piezoelectric materials (**A**), leading to the development of the CrystalDFT database (**Figure S2**). The panels present a series of graphs, each highlighting structural characteristics (**B** and **C**), crystal systems (**D**), point groups (**E**) and space group (**F**) information within the CrystalDFT dataset.

*High-throughput Workflow Overview:* Preparing, managing and analyzing the results of DFT calculations is a time-consuming task for small batches of crystals, but rapidly becomes intractable for large batches of tens or hundreds. High-throughput computational methods can minimise the human time required for batch calculations, utilizing automation to streamline and parallelize each stage of the calculation process (**Figure 7**). There are three stages of completing a computational calculation of a material's characteristics using VASP: File Preparation, Calculation Submission and Maintenance, and Output Analysis. Sequential scripts were developed for each of these stages.

*File Preparation:* File preparation consists of creating a variety of files and folders to produce a submission ready calculation set. VASP takes a variety of text files as inputs, so much of the file preparation script consists of python code to create, populate, and move text files. The script begins by converting all provided CIF files into POSCAR files, the crystal structure file format required for VASP calculations. This involves an initial conversion performed using the Atomic Simulation Environment python module[49], followed by a series of formatting stages which ensure the VASP compatibility of the file. Once this primary step is completed, a series of text files are generated which specify the conditions of the calculation (POTCAR, KPOINTS, INCAR, sub.sh), as well as a series of scripts used for calculation, submission, and maintenance (sub.sh, Automate.sh) The resulting calculation folder is uploaded to the supercomputer being used, and the script initializes the initial structural relaxation calculations for the entire batch.

*Calculation Submission and monitoring:* The process of monitoring calculations and submitting new ones when necessary is a lengthy and tedious process when done manually. It involves repeatedly checking ongoing calculations by logging onto a remote server and waiting hours or days for emails from scheduling software. This is reasonably straightforward when the number of crystals being characterized is small, but rapidly becomes intractable for larger batches. There are a minimum of three sequential calculations required for each characterisation in this work, with calculations for different materials in a batch being run asynchronously due to computing resource constraints. To maximize group calculation speed, each crystal should be managed separately, with further calculations being initialized as soon as the prerequisites have been completed. This task is completed by the monitoring script, which is launched each time a calculation finishes. Once launched, it rapidly searches the job queue and uses a decision tree derived from the manual process to determine the stage that each crystal is in, submitting any required calculations. Following the final calculation in the batch, the output analysis script is initialized.

*Output Analysis Script:* A script was developed to parse through the OUTCAR files from the finished DFT calculations, extracting important information about the properties of the crystals being characterized such as the Dielectric, Elastic and Piezoelectric tensors. Once analyzed, the resulting data is screened for unstable behaviour using a variety of flags which indicate the validity of the computed properties. Crystals can be flagged for a variety of reasons such as high dielectric constants and elastic tensors with either negative eigenvalues, or large condition numbers. This screening identifies faulty crystals which have concerning properties that might indicate either low-quality input CIF files, or calculation errors. In the context of large-scale database creation, these checks provide an extremely important quality control element to the material characterisation process which identifies and flags any materials which may not be suitable for input into the database (**Figure 7**).

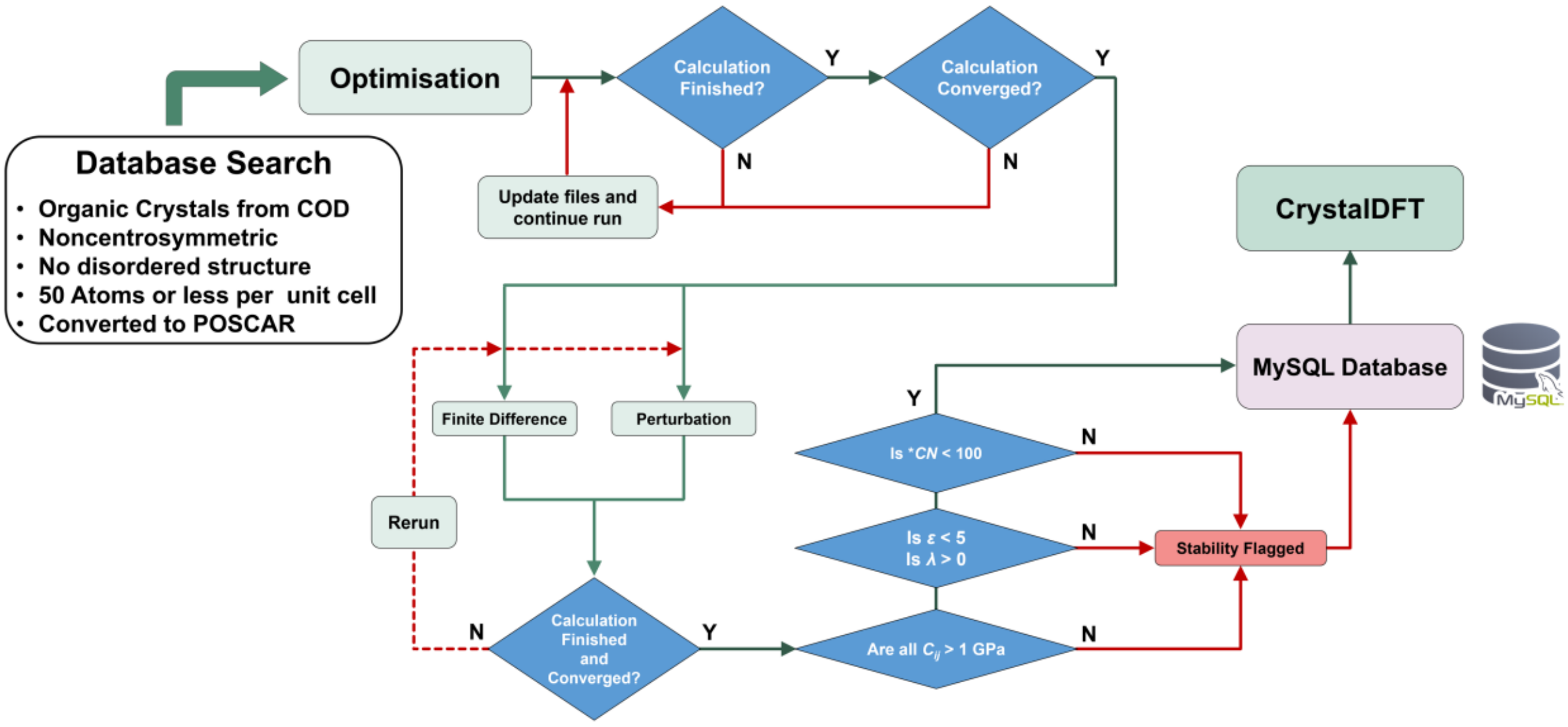

**Figure 7.** High-throughput computational screening workflow for adding crystal structures to the CrystalDFT database with predicted piezoelectric constants. Starting with initial non-centrosymmetric organic crystal structures from COD database, the process progresses through preprocessing, DFT setup and execution, piezoelectric property calculations, and rigorous data validation. Each stage incorporates error-checking to ensure data integrity. Finalised, validated results, including piezoelectric constants and structural parameters, are then stored in the CrystalDFT database, completing the workflow. All flag criteria are detailed in **Supplementary Note S1**.

*DFT Parameters:* All calculations were performed using periodic DFT[50] with the *Vienna Ab initio Simulation Package (VASP)*[51] code. To predict the electronic structures, we used the PBE functional[52] with Grimme-D3 dispersion corrections[53] and projector augmented wave (PAW) pseudopotentials[54] with a plane wave cut-off of 800 eV and k-point sampling of 4x4x4. A finite differences method was used to calculate the stiffness tensor, with each atom being displaced in each direction by ±0.01 Å (plane wave cut-off of 600 eV, and k-point sampling of 2×2×2). Within VASP, the elastic tensors were determined through the finite difference method. This method entails perturbing the crystal lattice along each lattice vector and computing the resulting polarization. The elastic constants are represented in the form of the stiffness tensor, C, as a 6×6 matrix. VASP outputs this tensor in Voigt notation in kB, so post-analysis was required to produce C in matrix notation in GPa. Piezoelectric strain constants and dielectric tensors were calculated using Density Functional Perturbation Theory[55] (DFPT) (plane wave cut-off of 600 eV, and k-point sampling of 2×2×2). Using the piezoelectric charge coefficients, $e_{ij}$, which are calculated directly by VASP code, and the elastic stiffness constants, $c_{kj}$, we derived the more useful piezoelectric strain coefficient, $d_{ik}$. This derived piezoelectric tensor establishes a connection between induced polarization and applied strain, facilitating the determination of piezoelectric coefficients in pC/N (or equivalent converse piezoelectric units pm/V). All piezoelectric responses are described by a third-rank tensor in the form of a 3×6 matrix (see **eq. 1** and the matrices (**eq. 2**) below).

$$d_{ik} = \frac{e_{ij}}{c_{kj}} \tag{1}$$

$$\begin{bmatrix} d_{11} & d_{12} & d_{13} & d_{14} & d_{15} & d_{16} \\ d_{21} & d_{22} & d_{23} & d_{24} & d_{25} & d_{26} \\ d_{31} & d_{32} & d_{33} & d_{34} & d_{35} & d_{36} \end{bmatrix} \tag{2}$$

The number of non-zero elements in both the elastic and piezoelectric matrices vary according to the symmetry of the crystal being studied[56]. We used our in-house code, which has multiple functionalities for pre- and post-processing of DFT simulations data including the analysis of elastic tensors. It takes a 6×6 tensor as an input to calculate various elastic properties, such as stiffness, compliance, young's moduli, shear moduli *etc.* The diagonal elements represent the normal or longitudinal stiffness coefficients. Whereas the off diagonal elements represent how different directions of stress and strain are connected to each other, indicating that the material behaves differently depending on the direction of the applied force. This means the material has different properties in different directions, which is known as anisotropy. Young's Moduli were derived from the stiffness and its inverse compliance matrix components and the values are represented as Voigt-Reuss-Hill averages were calculated using our in-house code. We used VESTA[57], a 3D visualisation program to visualise crystal structures.

*Functional Group Analysis:* Functional group analysis was conducted using a combination of in-house scripts[58] and established cheminformatics tools. SMILES notation for each crystal molecular entity was generated using Open Babel[59], and RDKit[60] was subsequently employed to identify and analyse the presence of specific functional groups. This approach allowed for accurate characterisation of molecular features within the CrystalDFT dataset, enhancing the reliability of functional group identification and piezoelectric property correlations.

**Database Interpretation: Piezoelectric Crystal Theory**

Crystals that lack inversion symmetry are the only ones with the potential to exhibit piezoelectricity. Out of the 32 crystallographic point groups, 21 are non-centrosymmetric, meaning they do not possess inversion symmetry (**Figure 8**). Of these, 20 point groups exhibit piezoelectric behaviour, with the exception of point group 432[61]. The piezoelectric point groups can be further classified into two categories: ten are polar, while the remaining ten are nonpolar. Interestingly, the vast majority of bioorganic materials naturally lack a center of symmetry in their crystal structures, making them inherently suited for piezoelectric applications. This lack of centrosymmetry in bio/organic compounds stems from the complex, asymmetric arrangements of molecular space within the crystals. As a result, these materials display significant potential for piezoelectricity, opening pathways for their integration into a range of technological applications, including sensors, actuators and energy harvesting devices.

Each space group in a crystal corresponds to a unique symmetry configuration. For our study, we are utilising 3×6 tensors to understand the piezoelectric coefficients. This helps us to understand how the space group exhibits the least amount of symmetry. If a crystal has high symmetry, many components of the tensor will be zero due to symmetry restrictions, reducing the crystal's overall piezoelectric response. On the other hand, lower-symmetry space groups allow more non-zero components in the piezoelectric tensor, increasing the potential piezoelectric effect. This is because fewer symmetry elements mean fewer constraints on the directional alignment of dipoles in response to stress. For instance, a highly asymmetric crystal may have a piezoelectric tensor with several non-zero components, indicating that it can produce an electric polarisation along multiple directions when stressed and since triclinic space group contains most of the non-zero coefficients and hence considered as the crystallography system with the lowest symmetry. Each space group corresponds to a distinct set of piezoelectric tensor components, as illustrated in **Figure 8.**

| *Crystal System* | *Noncentrosymmetric Point Groups* | *Piezoelectric Strain Tensors* |
|---|---|---|
| *Triclinic* | 1 | $\begin{pmatrix} d_{11} & d_{12} & d_{13} & d_{14} & d_{15} & d_{16} \\ d_{21} & d_{22} & d_{23} & d_{24} & d_{25} & d_{26} \\ d_{31} & d_{32} & d_{33} & d_{34} & d_{35} & d_{36} \end{pmatrix}$ |
| *Monoclinic* | 2, *m* | $\begin{pmatrix} 0 & 0 & 0 & d_{14} & 0 & d_{16} \\ d_{21} & d_{22} & d_{23} & 0 & d_{25} & 0 \\ 0 & 0 & 0 & d_{34} & 0 & d_{36} \end{pmatrix}$ $\begin{pmatrix} d_{11} & d_{12} & d_{13} & 0 & d_{15} & 0 \\ 0 & 0 & 0 & d_{24} & 0 & d_{26} \\ d_{31} & d_{32} & d_{33} & 0 & d_{35} & 0 \end{pmatrix}$ |
| *Orthorhombic* | 222, *mm*2 | $\begin{pmatrix} 0 & 0 & 0 & d_{14} & 0 & 0 \\ 0 & 0 & 0 & 0 & d_{25} & 0 \\ 0 & 0 & 0 & 0 & 0 & d_{36} \end{pmatrix}$ $\begin{pmatrix} 0 & 0 & 0 & 0 & d_{15} & 0 \\ 0 & 0 & 0 & d_{24} & 0 & 0 \\ d_{31} & d_{32} & d_{33} & 0 & 0 & 0 \end{pmatrix}$ |
| *Tetragonal* | 4, $\bar{4}$, 4mm, $\bar{4}2m$ | $\begin{pmatrix} 0 & 0 & 0 & d_{14} & d_{15} & 0 \\ 0 & 0 & 0 & d_{15} & -d_{14} & 0 \\ d_{31} & d_{31} & d_{33} & 0 & 0 & 0 \end{pmatrix}$ $\begin{pmatrix} 0 & 0 & 0 & d_{14} & d_{15} & 0 \\ 0 & 0 & 0 & -d_{15} & d_{14} & 0 \\ d_{31} & -d_{31} & 0 & 0 & 0 & d_{36} \end{pmatrix}$ $\begin{pmatrix} 0 & 0 & 0 & 0 & d_{15} & 0 \\ 0 & 0 & 0 & d_{15} & 0 & 0 \\ d_{31} & d_{31} & d_{33} & 0 & 0 & 0 \end{pmatrix}$ $\begin{pmatrix} 0 & 0 & 0 & d_{14} & 0 & 0 \\ 0 & 0 & 0 & 0 & d_{14} & 0 \\ 0 & 0 & 0 & 0 & 0 & d_{36} \end{pmatrix}$ |
| *Trigonal* | 3, 32, 3*m* | $\begin{pmatrix} d_{11} & -d_{11} & 0 & d_{14} & d_{15} & -d_{22} \\ -d_{22} & d_{22} & 0 & d_{15} & -d_{14} & -d_{11} \\ d_{31} & d_{31} & d_{33} & 0 & 0 & 0 \end{pmatrix}$ $\begin{pmatrix} d_{11} & -d_{11} & 0 & d_{14} & 0 & 0 \\ 0 & 0 & 0 & 0 & -d_{14} & -d_{11} \\ 0 & 0 & 0 & 0 & 0 & 0 \end{pmatrix}$ $\begin{pmatrix} 0 & 0 & 0 & 0 & d_{15} & -d_{22} \\ -d_{22} & d_{22} & 0 & d_{15} & 0 & 0 \\ d_{31} & d_{32} & d_{33} & 0 & 0 & 0 \end{pmatrix}$ |
| *Hexagonal* | 6, $\bar{6}$, 6*mm*, $\bar{6}m2$ | $\begin{pmatrix} 0 & 0 & 0 & d_{14} & d_{15} & 0 \\ 0 & 0 & 0 & d_{15} & -d_{14} & 0 \\ d_{31} & d_{31} & d_{33} & 0 & 0 & 0 \end{pmatrix}$ $\begin{pmatrix} d_{11} & -d_{11} & 0 & 0 & 0 & -d_{22} \\ -d_{22} & d_{22} & 0 & 0 & 0 & -d_{11} \\ 0 & 0 & 0 & 0 & 0 & 0 \end{pmatrix}$ $\begin{pmatrix} 0 & 0 & 0 & 0 & d_{15} & 0 \\ 0 & 0 & 0 & d_{15} & 0 & 0 \\ d_{31} & d_{31} & d_{33} & 0 & 0 & 0 \end{pmatrix}$ $\begin{pmatrix} 0 & 0 & 0 & 0 & 0 & -d_{22} \\ -d_{22} & d_{22} & 0 & 0 & 0 & 0 \\ 0 & 0 & 0 & 0 & 0 & 0 \end{pmatrix}$ |
| *Cubic* | 23, $\bar{4}3m$ | $\begin{pmatrix} 0 & 0 & 0 & d_{14} & 0 & 0 \\ 0 & 0 & 0 & 0 & d_{14} & 0 \\ 0 & 0 & 0 & 0 & 0 & d_{14} \end{pmatrix}$ |

**Figure 8.** Piezoelectric strain coefficient tensors for non-centrosymmetric point groups in different crystal system types. The matrices of piezoelectric coefficients are arranged according to the sequence of point groups in the second column, with only 18 non-centrosymmetric point groups listed out of 21. Piezoelectric point groups 422 and 622 are excluded, as no electric charge is generated under applied compression; these groups display only transverse compressional or torsional piezoelectric effects. Additionally, in the non-centrosymmetric point group 432, all piezoelectric moduli are zero due to its high symmetry.

*Elastic Tensors analysis:* This tensor helps in understanding the directional dependence of the material's elastic stiffness and flexibility. Organic crystals are generally very flexible, while still producing excellent piezoelectric responses. This is due to the fact that individual molecules in the crystal lattice are held together by supramolecular forces such as hydrogen bonds which are weak in comparison to the covalent and ionic bonds present in traditional inorganic piezoelectric materials. These stronger bonds result in inorganic materials like PZT being much more rigid than organic crystals. Our database also includes 2D and 3D representations of the spatial dependence of Young’s modulus and shear modulus, allowing users to observe the directional dependence of elasticity.

*Strain Tensor analysis:* This is a 3×6 tensor, where the diagonal elements ($d_{11}$, $d_{22}$, $d_{33}$) represent the longitudinal strain and $d_{44}$, $d_{55}$, $d_{66}$ represents the shear strain.

## Validation of computational predictions against experimental data

To assess the reliability of the calculated piezoelectric constants, we conducted a thorough comparison with available experimental data. This evaluation encompassed 16 reported single crystal systems of bioorganic and inorganic materials and involved 30 distinct components of the piezoelectric strain tensors. The systems selected for this analysis are detailed in **Table S4** and **Table S5**. The results of this comparison between the calculated and experimental values for the piezoelectric constants are presented in **Figure 9A**. This rigorous validation process not only highlights the accuracy of our computational high-throughput simulations pipeline but also underscores their potential for predicting the piezoelectric properties of bioorganic materials.

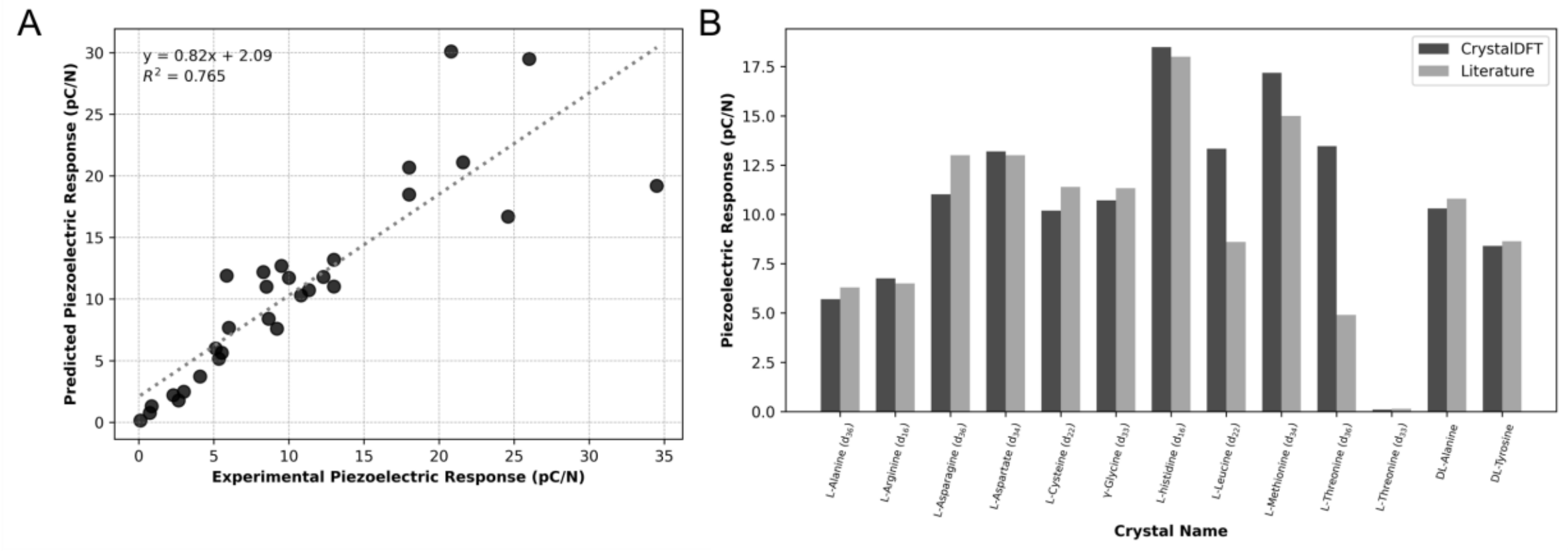

**Figure 9.** Validation of the computational methodology for predicting piezoelectric constants. (a) Regression plot comparing calculated piezoelectric constants with experimental data reported in the literature (**Table S4** and **Table S5**). The strong correlation ($R^2 = 0.76$) demonstrates a high level of agreement between computational predictions and literature values, validating the accuracy of the simulation pipeline. To enhance visualisation clarity, coefficients exceeding 40 pC/N were excluded from the plot. (b) Comparison of computational predictions with literature-reported experimental and predicted piezoelectric strain constants for amino acids, with detailed data provided in Supplementary **Table S6**.

The comparison of calculated and experimental piezoelectric constants for various bioorganic systems highlights strong correlations across distinct tensor components. For example, γ-Glycine (COD ID: 7128793) exhibited experimental strain coefficients of 5.33 pC/N ($d_{16}$) and 11.33 pC/N ($d_{33}$) from the literature[1], compared to DFT-predicted values of 5.15 pC/N and 10.72 pC/N, respectively. L-Histidine also showed a close match, with reported values of 18 pC/N ($d_{24}$) against predicted values of 18.49 pC/N and 20.68 pC/N for its two distinct entries, COD IDs 2108877 and 2108883. Other materials, such as L-Aspartate and DL-Alanine, displayed good agreement between literature values and predictions, further reinforcing the accuracy of the computational high-throughput simulations pipeline for predicting piezoelectric properties in bioorganic materials (**Figure 9B**). This analysis underscores the reliability of these simulations in elucidating the piezoelectric characteristics of various compounds.

Our database also encompasses a diverse range of bio/organic materials, some of which have been previously studied for their energetic profiles, piezoelectric, and optical properties (**Table S5**). This includes L-ornithine hydrochloride, reported for its nonlinear optical properties (NLO), and DL-malic acid (5000188), noted for its piezoelectric response. The derivatives of both malic acid and tartaric acid exhibit significant piezoelectric characteristics, with the latter showing a maximum predicted strain of -21.47. L-threonine has demonstrated a predicted $d_{33}$ value of 0.14, while N-methylurea has been examined in related studies. Graphene oxide has been investigated both computationally and experimentally, confirming its contribution to piezoelectricity. Additionally, methylpiperazine has been reported in chiral hybrid metal halides, exhibiting a $d_{14}$ value of -10.58, while cristobalite has related studies available. Furthermore, 5-aminotetrazolium dinitramide is recognised for its energetic material derivatives, and p-nitrobenzaldehyde has been documented for its electrical, optical, and mechanical properties. This comprehensive database highlights the potential of these organic materials for future advancements in piezoelectric applications.

## Associated Content

**Data availability.** All data presented in this study, along with the necessary resources to reproduce the findings, are available on the official CrystalDFT database website at https://actuatelab.ie/CrystalDFT/index.html.

**Code availability**. All the source code developed for this work is available on Actuate Lab website at https://actuatelab.ie/actuate_tools.html.

**Supplementary information**. SI contains supplementary notes S1 to S2, figures S1 to S2 and tables S1 to S6.

## Corresponding author

Correspondence to Sarah Guerin. E-mail: Sarah.Guerin@ul.ie

## Competing interests

The authors declare no competing interests.

## Acknowledgments

S.V., G.K., and S.G. are funded by the European Union under ERC Starting Grant no. 101039636. S. G. would like to acknowledge funding from Research Ireland under grant number 21/PATH-S/9737, and R.G. is funded via RI grant 12/RC/2275_P2. All authors acknowledge ongoing support from the Irish Centre for High-End Computing (ICHEC).

**Supporting Information**

# High-throughput computational screening of small, eco-friendly, molecular crystals for sustainable piezoelectric materials

Shubham Vishnoi[1], Geetu Kumari[1] , Robert Guest[1,2], Pierre-André Cazade[3*],

& Sarah Guerin[1, 2*]

[1]*Department of Chemical Sciences, Bernal Institute, University of Limerick, V94 T9PX, Ireland*

[2]*SSPC, The Science Foundation Ireland Research Centre for Pharmaceuticals, University of Limerick, V94 T9PX, Ireland*

[3]*Department of Physics, Bernal Institute, University of Limerick, V94 T9PX, Ireland*

# Table of Contents

***Supplementary Notes***

***List of Figures***

***List of Tables***

***Supplementary Notes***

***S1. Flagging Criteria in CrystalDFT High-throughput Workflow:*** *In the high-throughput workflow described in this study, a set of criteria is used to flag crystal structures for further examination when predicted properties fall outside predefined ranges. These criteria help ensure that the workflow focuses on stable, meaningful results and filters out anomalous outputs caused by errors or instabilities. Below is a detailed explanation of the flagging process and its implications:*

***1. Abnormal Dielectric Constant:*** *Crystals with a* ***predicted dielectric constant*** *(ε) greater than* ***5*** *are flagged. Such values often indicate protonation errors or other inaccuracies in the initial structure or calculations.*

***2. Mechanical Instability:*** *Crystals with a predicted elastic constant below* ***1 GPa*** *are flagged. This suggests structural instability, as such low stiffness values are inconsistent with mechanically sound materials.*

***3. Numerical Instabilities:*** *Crystals with* ***high condition numbers*** *(CN > 100) are flagged. A high condition number implies numerical instability in the calculations, which can lead to unreliable results.*

***4. Negative Eigenvalues:*** *Crystals with* ***negative eigenvalues*** *in the calculated tensors are flagged. This typically indicates an issue with the stability of the structure or calculation errors[1].*

***S2. Workflow Overview:*** *The automated workflow integrates these criteria into its decision-making process. Structures that meet the above conditions are automatically excluded from further high-throughput screening and marked with flagged notation. *However, the condition number flag is not applied stringently across all analyses and is only used for specific figures, as noted in the captions for those particular figures. All other flags are applied rigorously throughout the study. This could prevent errors from propagating through the dataset and allows future studies and computational resources to focus on promising candidates.*

*CrystalDFT Workflow Script and Database: The workflow was constructed using Python-based libraries, which facilitates the pre-selection, input file preparation (Atomic Simulation Environment, ASE[2]), formatting, optimization (via DFT - VASP[3]), electromechanical property prediction of the input crystal structures and output processing (for additional details and access to scripts, please refer to - https://actuatelab.ie/actuate_tools.html). The CrystalDFT database (**Figure S2 - https://actuatelab.ie/CrystalDFT/index.html**) also includes visualisation of tensors based on*

*calculation outputs, along with several widgets (such as HD Viewer, 3D Visualiser – integrated with JSmol*[4]*, Crystal Card), illustrating key details related to crystal structure (including symmetry and molecular information), predicted tensors, and decision points such as flagging and further analysis. We used MTEX*[5] *for visualising polarisation in 3D. This rigorous flagging and evaluation process ensures high reliability and efficiency in identifying promising organic piezoelectric materials.*

*Testing and Screening Outcomes: A total of* ***572 crystal structures*** *were evaluated in this study. Structures flagged for any of the above criteria were excluded from the final analysis. After applying these filters, 22 crystals were identified that exhibited double-digit piezoelectric coefficients (measured in pC/N), meeting the stringent stability and performance thresholds defined in the workflow.*

## *Figures*

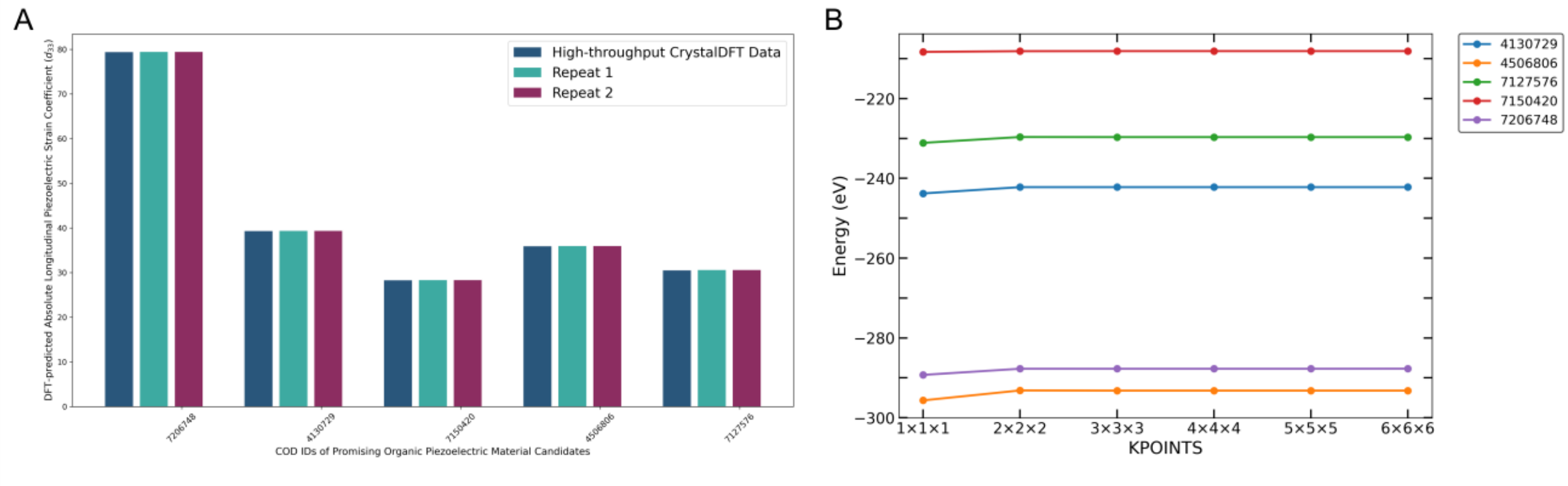

**Figure S1.** Validation of promising CrystalDFT-screened organic piezoelectric candidates: **(A)** Predicted $d_{33}$ values validated by repeated DFT calculations, **(B)** Total energy difference *versus* K-point convergence curves for the most promising organic piezoelectric materials identified during the screening process.

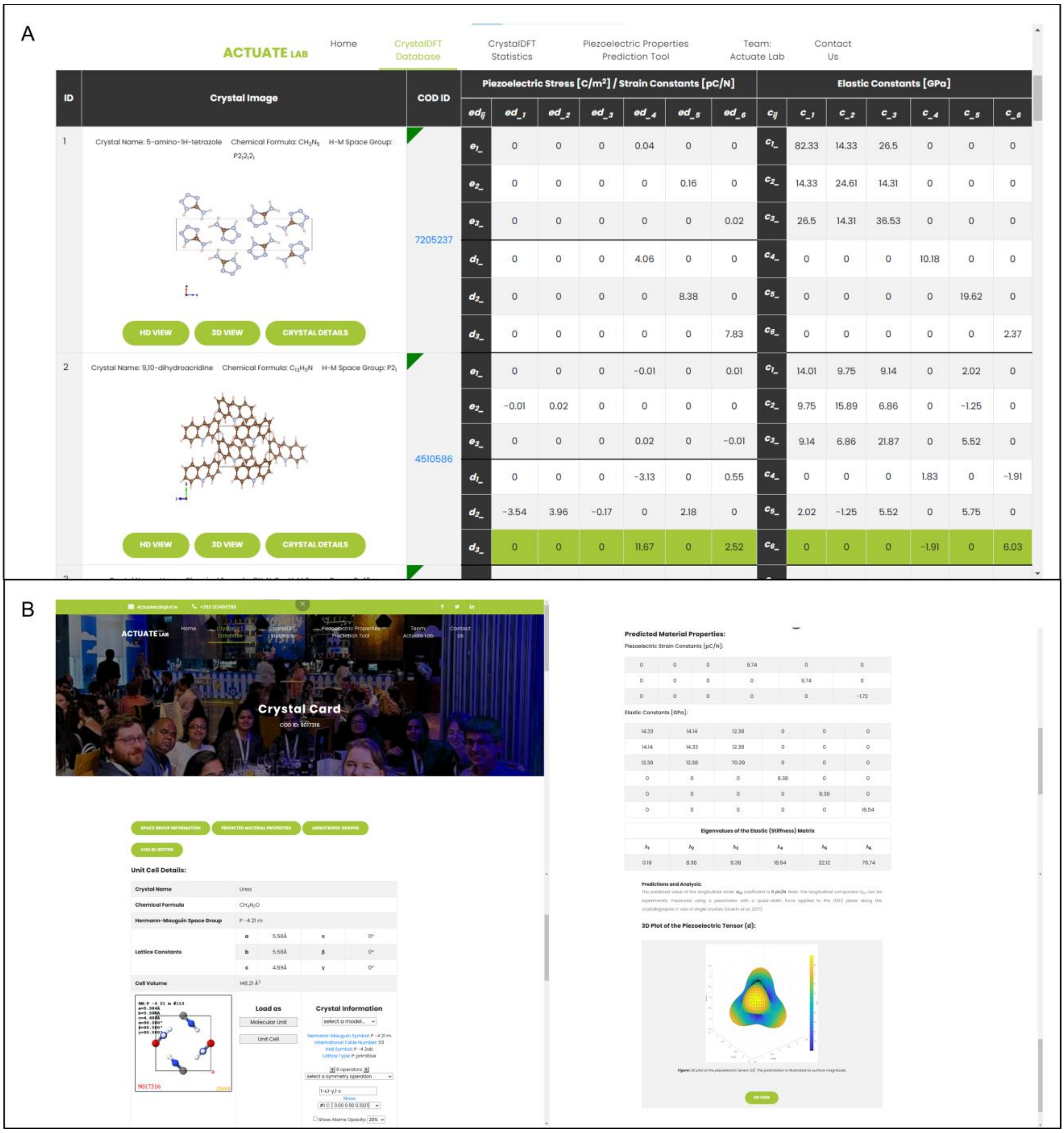

**Figure S2.** The CrystalDFT database provides a comprehensive set of tools for visualising and analysing our DFT-predicted piezoelectric tensors and crystal data along with the links to the relevant resources. **(A)** It includes advanced visualisation features for tensors derived from calculation outputs, along with interactive widgets such as the HD Viewer, 3D Visualiser, and **(B)** Crystal Card. These tools offer detailed insights into crystal structures, including symmetry and molecular information, as well as predicted tensors and key decision points, such as flagging details and other derived properties. Database Website Link: https://actuatelab.ie/CrystalDFT/index.html.

Table S**1A**. Reported maximum piezoelectric response in pC/N for a variety of piezoelectric materials, as plotted in Figure 1 in the main text.

| **Material** | **Piezoelectric Strain Coefficient (pC/N)** | **Category** |
|---|---|---|
| Collagen[6] | 12 | Bioorganics |
| γ-Glycine[7] | 9.9 | Bioorganics |
| L-Alanine[8] | 6 | Bioorganics |
| Hydroxy-L-Proline (Single Crystal)[9] | 25 | Bioorganics |
| Hydroxy-L-Proline (Film)[10] | 1.5 | Bioorganics |
| L-Threonine (Film)[10] | 0.2 | Bioorganics |
| Hydroxyapatite (Poled Ceramic)[11] | 2 | Bioorganics |
| M13 Bacteriophage Virus[12] | 10 | Bioorganics |
| Bone[13] | 0.2 | Bioorganics |
| Wood[14] | 0.1 | Bioorganics |
| Dry Tendon[15] | 2 | Bioorganics |
| Wool[16] | 0.1 | Bioorganics |
| Horn[16] | 1.8 | Bioorganics |
| Starch[16] | 2 | Bioorganics |
| Glycine Nanofibers[17] | 12.5 | Bioorganics |
| Spar[18] | 0.02 | Bioorganics |
| Lysozyme (Film)[19] | 6.5 | Bioorganics |
| Chitin Nanofiber[20] | 2 | Bioorganics |
| Gelatine Nanofiber[21] | 20 | Bioorganics |
| Di-Phenylalanine (FF) Peptide Nanotubes[22] | 60 | Bioorganics |
| Chitosan[23] | 6 | Bioorganics |
| CNF[23] | 8 | Bioorganics |
| Eggshell Membrane[24] | 18.3 | Bioorganics |
| Silk[25] | 56.2 | Bioorganics |
| Di-Phenylalanine (FF) Microribbon[26] | 40 | Bioorganics |
| L-Asparagine[7] | 13 | Bioorganics |
| L-Leucine[7] | 20 | Bioorganics |
| L-Histidine[7] | 18 | Bioorganics |
| L-Methionine[7] | 15 | Bioorganics |
| L-Aspartate[7] | 13 | Bioorganics |
| L-Isoleucine[7] | 25 | Bioorganics |
| L-Cysteine[7] | 11 | Bioorganics |
| DL-Alanine[27] | 4 | Bioorganics |
| Poly(Î³-benzyl-l-glutamate)(PBLG)[28] | 25 | Biopolymers |
| Polymethyl-L-Glutamate (PMG)[16] | 2 | Biopolymers |
| Poly-Lactic-Acid (PLA)[16] | 10 | Biopolymers |
| Porous Polypropylene (p-PP)[29] | 450 | Polymers |
| Poly-b-Hydroxybutyrate[16] | 1.3 | Polymers |
| PVDF (Poled)[30] | 32 | Polymers |
| Propylene Oxide[16] | 0.1 | Polymers |
| P(VDF-TRFE)[31] | 40 | Polymers |
| Lithium Niobate[32] | 42 | Inorganics |
| Quartz[33] | 2.3 | Inorganics |
| $BiB_3O_6$[34] | 40 | Inorganics |
| $BiFeO_3$[35] | 18 | Inorganics |
| $BaTiO_3$ | 90 | Inorganics |
| $Al_3AsO_7$[36] | 9 | Inorganics |
| $Al_3PO_7$[36] | 3.4 | Inorganics |
| $Ga_3AsO_7$[36] | 23.9 | Inorganics |
| $GaPO_7$[36] | 8.2 | Inorganics |
| ZnO[37] | 9.9 | Inorganics |
| AlN[38] | 5.4 | Inorganics |
| Barium Titanate (Single Crystal)[39] | 392 | Inorganics |
| Rochelle Salt[40] | 345 | Inorganics |
| CBT[41] | 8 | Inorganics |
| NBBT95/5[42] | 360 | Inorganics |
| Mn-KNN[43] | 350 | Inorganics |
| BZT-BCT[44] | 637 | Inorganics |
| Lead Zirconium Titanate (PZT)[45] | 493 | Lead-containing |
| Barium Titanate (Ceramic)[39] | 270 | Inorganics |
| Potassium-Sodium Niobate[46] | 490 | Inorganics |
| $GaPBi_3$[47] | 146 | Inorganics |
| GeSe[48] | 212 | Inorganics |
| SnS[48] | 145 | Inorganics |
| SnSe[48] | 251 | Inorganics |
| GeS[48] | 75 | Inorganics |
| ZnO Nanorods[49] | 9.5 | Nanostructures |
| PZT Nanoshell[50] | 90 | Lead-containing |
| GaN Nanowires[51] | 12.8 | Nanostructures |
| PZT ceramic[52] | 407.4 | Lead-containing |

**Table 1B.** Summary of the averages and ranges for each category listed in Supplementary Information **Table S1A**, with the median providing an additional measure of central tendency. These values indicate currently observed potential for applications in energy harvesting and sensing, with detailed data in **Table S1A**.

| Category | Average (pC/N) | Median (pC/N) | Range | Standard Deviation |
|---|---|---|---|---|
| **Bioorganics** | 13.11 | 10 | 0.02 to 60 | 15.28 |
| **Biopolymers** | 12.33 | 10 | 2 to 25 | 11.68 |
| **Polymers** | 104.68 | 32 | 0.1 to 450 | 193.87 |
| **Inorganics** | 167.09 | 75 | 2.3 to 637 | 186.10 |
| **Lead-containing** | 330.13 | 407.4 | 90 to 493 | 212.32 |
| **Nanostructures** | 11.15 | 11.15 | 9.5 to 12.8 | 2.33 |

**Table S2.** Summary of functional group representation within the CrystalDFT database, showing total occurrences of each functional group and their respective percentage within the dataset. The table also includes the count of each functional group within the HTS unflagged subset of CrystalDFT database and the percentage representation of each within this subset. Notably, we used RDKit[53] to analyse functional groups, following the generation of SMILES notation for each crystal molecular entity using Open Babel[54]. Crystals for which RDKit analysis failed were excluded from the analysis. This subset is refined to exclude crystals with high dielectric constants, negative stiffness or high condition numbers, ensuring a focus on functional groups that contribute to stable and reliable piezoelectric properties.

| **Functional Group** | **Full CrystalDFT Dataset** | | **HTS Unflagged Subset** | |
|---|---|---|---|---|
| | Total Count | Percentage (%) | Total Count | Percentage (%) |
| Hydroxyl | 216 | 49 | 145 | 57.8 |
| Aldehyde | 257 | 58.3 | 151 | 60.2 |
| Ketone | 148 | 33.6 | 102 | 40.6 |
| Carboxylic Acid | 140 | 31.7 | 90 | 35.9 |
| Amino | 263 | 59.6 | 151 | 60.2 |
| Amide | 128 | 29 | 79 | 31.5 |
| Ester | 140 | 31.7 | 90 | 35.9 |
| Nitrile | 26 | 5.9 | 15 | 6 |
| Alkane | 368 | 83.4 | 220 | 87.6 |
| Alkene | 73 | 16.6 | 37 | 14.7 |
| Alkyne | 9 | 2 | 5 | 2 |
| Ether | 85 | 19.3 | 70 | 27.9 |
| Cycloalkane | 17 | 3.9 | 10 | 4 |
| Aromatic Ring | 148 | 33.6 | 90 | 35.9 |
| Pyridine | 14 | 3.2 | 7 | 2.8 |

**Table S3.** Hermann-Mauguin Space Group Symbols and Corresponding Space Group Numbers. Only structures with space groups 1, 3–9, 16–46, 75–82, 89–122, 143–146, 149–161, 168–174, 177–190, 195–199, 207–220 are allowed (since these space groups lack inversion symmetry).

| | | | | | | | | | |
|---|---|---|---|---|---|---|---|---|---|
| 1 | P 1 | 2 | P -1 | 3 | P 2 | 4 | $P\ 2_1$ | 5 | C 2 |
| 6 | P m | 7 | P c | 8 | C m | 9 | C c | 10 | P 2/m |
| 11 | $P\ 2_1/m$ | 12 | C 2/m | 13 | P 2/c | 14 | $P\ 2_1/c$ | 15 | C 2/c |
| 16 | P 2 2 2 | 17 | $P\ 2\ 2\ 2_1$ | 18 | $P\ 2_1\ 2_1\ 2$ | 19 | $P\ 2_1\ 2_1\ 2_1$ | 20 | $C\ 2\ 2\ 2_1$ |
| 21 | C 2 2 2 | 22 | F 2 2 2 | 23 | I 2 2 2 | 24 | $I\ 2_1\ 2_1\ 2_1$ | 25 | P m m 2 |
| 26 | $P\ m\ c\ 2_1$ | 27 | P c c 2 | 28 | P m a 2 | 29 | $P\ c\ a\ 2_1$ | 30 | P n c 2 |
| 31 | P m n $2_1$ | 32 | P b a 2 | 33 | P n a $2_1$ | 34 | P n n 2 | 35 | C m m 2 |
| 36 | C m c $2_1$ | 37 | C c c 2 | 38 | A m m 2 | 39 | A b m 2 | 40 | A m a 2 |
| 41 | A b a 2 | 42 | F m m 2 | 43 | F d d 2 | 44 | I m m 2 | 45 | I b a 2 |
| 46 | I m a 2 | 47 | P m m m | 48 | P n n n | 49 | P c c m | 50 | P b a n |
| 51 | P m m a | 52 | P n n a | 53 | P m n a | 54 | P c c a | 55 | P b a m |
| 56 | P c c n | 57 | P b c m | 58 | P n n m | 59 | P m m n | 60 | P b c n |
| 61 | P b c a | 62 | P n m a | 63 | C m c m | 64 | C m c a | 65 | C m m m |
| 66 | C c c m | 67 | C m m a | 68 | C c c a | 69 | F m m m | 70 | F d d d |
| 71 | I m m m | 72 | I b a m | 73 | I b c a | 74 | I m m a | 75 | P 4 |
| 76 | P $4_1$ | 77 | P $4_2$ | 78 | P $4_3$ | 79 | I 4 | 80 | I $4_1$ |
| 81 | P -4 | 82 | I -4 | 83 | P 4/m | 84 | P $4_2$/m | 85 | P 4/n |
| 86 | P $4_2$/n | 87 | I 4/m | 88 | I $4_1$/a | 89 | P 4 2 2 | 90 | P 4 $2_1$ 2 |
| 91 | P $4_1$ 2 2 | 92 | P $4_1$ $2_1$ 2 | 93 | P $4_2$ 2 2 | 94 | P $4_2$ $2_1$ 2 | 95 | P $4_3$ 2 2 |
| 96 | P $4_3$ $2_1$ 2 | 97 | I 4 2 2 | 98 | I $4_1$ 2 2 | 99 | P 4 m m | 100 | P 4 b m |
| 101 | P $4_2$ c m | 102 | P $4_2$ n m | 103 | P 4 c c | 104 | P 4 n c | 105 | P $4_2$ m c |
| 106 | P $4_2$ b c | 107 | I 4 m m | 108 | I 4 c m | 109 | I $4_1$ m d | 110 | I $4_1$ c d |
| 111 | P -4 2 m | 112 | P -4 2 c | 113 | P -4 $2_1$ m | 114 | P -4 $2_1$ c | 115 | P -4 m 2 |
| 116 | P -4 c 2 | 117 | P -4 b 2 | 118 | P -4 n 2 | 119 | I -4 m 2 | 120 | I -4 c 2 |
| 121 | I -4 2 m | 122 | I -4 2 d | 123 | P 4/m m m | 124 | P 4/m c c | 125 | P 4/n b m |
| 126 | P 4/n n c | 127 | P 4/m b m | 128 | P 4/m n c | 129 | P 4/n m m | 130 | P 4/n c c |
| 131 | P $4_2$/m m c | 132 | P $4_2$/m c m | 133 | P $4_2$/n b c | 134 | P $4_2$/n n m | 135 | P $4_2$/m b c |
| 136 | P $4_2$/m n m | 137 | P $4_2$/n m c | 138 | P $4_2$/n c m | 139 | I 4/m m m | 140 | I 4/m c m |
| 141 | I $4_1$/a m d | 142 | I $4_1$/a c d | 143 | P 3 | 144 | P $3_1$ | 145 | P $3_2$ |
| 146 | R 3 | 147 | P -3 | 148 | R -3 | 149 | P 3 1 2 | 150 | P 3 2 1 |
| 151 | P $3_1$ 1 2 | 152 | P $3_1$ 2 1 | 153 | P $3_2$ 1 2 | 154 | P $3_2$ 2 1 | 155 | R 3 2 |
| 156 | P 3 m 1 | 157 | P 3 1 m | 158 | P 3 c 1 | 159 | P 3 1 c | 160 | R 3 m |
| 161 | R 3 c | 162 | P -3 1 m | 163 | P -3 1 c | 164 | P -3 m 1 | 165 | P -3 c 1 |
| 166 | R -3 m | 167 | R -3 c | 168 | P 6 | 169 | P $6_1$ | 170 | P $6_5$ |
| 171 | P $6_2$ | 172 | P $6_4$ | 173 | P $6_3$ | 174 | P -6 | 175 | P 6/m |
| 176 | P $6_3$/m | 177 | P 6 2 2 | 178 | P $6_1$ 2 2 | 179 | P $6_5$ 2 2 | 180 | P $6_2$ 2 2 |
| 181 | P $6_4$ 2 2 | 182 | P $6_3$ 2 2 | 183 | P 6 m m | 184 | P 6 c c | 185 | P $6_3$ c m |
| 186 | P $6_3$ m c | 187 | P -6 m 2 | 188 | P -6 c 2 | 189 | P -6 2 m | 190 | P -6 2 c |
| 191 | P 6/m m m | 192 | P 6/m c c | 193 | P $6_3$/m c m | 194 | P $6_3$/m m c | 195 | P 2 3 |
| 196 | F 2 3 | 197 | I 2 3 | 198 | P $2_1$ 3 | 199 | I $2_1$ 3 | 200 | P m -3 |
| 201 | P n -3 | 202 | F m -3 | 203 | F d -3 | 204 | I m -3 | 205 | P a -3 |
| 206 | I a -3 | 207 | P 4 3 2 | 208 | P $4_2$ 3 2 | 209 | F 4 3 2 | 210 | F $4_1$ 3 2 |
| 211 | I 4 3 2 | 212 | P $4_3$ 3 2 | 213 | P $4_1$ 3 2 | 214 | I $4_1$ 3 2 | 215 | P -4 3 m |
| 216 | F -4 3 m | 217 | I -4 3 m | 218 | P -4 3 n | 219 | F -4 3 c | 220 | I -4 3 d |
| 221 | P m -3 m | 222 | P n -3 n | 223 | P m -3 n | 224 | P n -3 m | 225 | F m -3 m |
| 226 | F m -3 c | 227 | F d -3 m | 228 | F d -3 c | 229 | I m -3 m | 230 | I a -3 d |

**Table S4.** Calculated piezoelectric strain coefficients for various well-known inorganic crystals, including zinc oxide, aluminium nitride, α-quartz and barium titanate, alongside experimental data for comprehensive comparison.

| Crystal Name (COD ID) | | Piezoelectric strain coefficients (pC/N) | | | | | |
|---|---|---|---|---|---|---|---|
| | | $d_{ij}$ | $d_{15}$ | $d_{31}$ | $d_{32}$ | $d_{33}$ | |
| **Zinc Oxide (ZnO)** | Predictions | | 12.2 | 6 | | 11.8 | |
| | Experiments[55] | | 8.3 | 5.1 | | 12.3 | |
| **Aluminium Nitride (AlN) (1605321)** | Predictions | | 3.7 | 1.7 | | 5.6 | |
| | Experiments[56,57] | | 4.08 | 2.65 | | 5.53 | |
| **α-quartz - Silicon Dioxide ($SiO_2$)** | Predictions | 2.2 ($d_{11}$) | 0.76 | | | | |
| | Experiments[58] | 2.3 ($d_{11}$) | 0.73 | | | | |
| **Barium Titanate ($BaTiO_3$)** | Predictions | | 2.5 | 19.2 | | 67.2 | |
| | Experiments[59] | | 392 | 34.5 | | 85.6 | 90 |
| **$KNbO_3$ (1594511)** | Predictions | 457 ($d_{24}$) | -2.7 | 16.7 | 12.7 | 21.1 | |
| | Experiments[60] | 241 ($d_{24}$) | 205 | 24.6 | 9.5 | 21.6 | |
| **Lithium Tantalate ($LiTaO_3$) (1594518)** | Predictions | 11 ($d_{22}$) | 29.5 | 2.48 | | 7.59 | |
| | Experiments[61] | 8.5 ($d_{22}$) | 26 | 3 | | 9.2 | |
| **Lithium Niobate ($LiNbO_3$)** | Predictions | 30.1 ($d_{22}$) | 93.1 | 1.31 | | 7.67 | |
| | Experiments[62] | 20.8 ($d_{22}$) | 69.2 | 0.85 | | 6 | |
| **Graphene oxide (1568395)** | | **Has been studied for its piezoelectric properties both computationally[63] and experimentally[64,65]** | | | | | |
| **Cristobalite (9015590)** | | **Related report available[66]** | | | | | |

**Table S5.** Summary of organic single crystal systems used for the comparison of calculated and experimental piezoelectric constants. For comparison purposes, we merged this data with the data from **Table S4**, which includes several well-known inorganic crystals and materials tested in-house, to provide a broader context against experimental measurements, as there are still very few small organic materials characterized for their piezoelectric properties. The table details the crystal name, reported experimental values, and corresponding calculated piezoelectric coefficients for each system.

| Crystal Name | COD ID/CSD ID | Experimental Piezoelectric strain coefficients (pC/N) | | | | | | DFT-predicted Piezoelectric strain coefficients (pC/N) | | | | | |
|---|---|---|---|---|---|---|---|---|---|---|---|---|---|
| | | $d_{11}$ | $d_{14}$ | $d_{15}$ | $d_{16}$ | $d_{31}$ | $d_{33}$ | $d_{11}$ | $d_{14}$ | $d_{15}$ | $d_{16}$ | $d_{31}$ | $d_{33}$ |
| γ-Glycine[7] | 7128793 | | | 5.33 | | | 11.33 | | | 5.15 | | | 10.72 |
| L-histidine | 2108877/2108883 | | | | | | 18 ($d_{24}$) | | | | | | 18.49/20.68 ($d_{24}$) |
| L-Aspartate | 2015971 | | | | 13 | | | | | | 13.2 | | |
| DL-Alanine | | | | | | | 10.8 | | | | | | 10.3 |
| DL-Tyrosine | | | | | | | 8.64 | | | | | | 8.4 |
| l-lactic acid | | | 10 | | | | | | 11.72 | | | | |
| β-Glycine | 4505645 | | | | 178 | | | | | | 50.66 | | 12.07 |
| Cellulose[67] | 4114383 | | | | | | 5.85 | | | | | | 11.91 |
| L-Asparagine | 2215317 | | | | | | 13 ($d_{16}$) | | | | | | 11.02 ($d_{16}$) |
| L-ornithine Hydrochloride | | Reported - NLO[68] | | | | | | | | | | | |
| DL-malic acid | 5000188 | Reported with Piezoelectric Response[69] | | | | | | | | | | | |
| Malic Acid | | Derivatives Reported with Piezoelectric Responses[70] | | | | | | | | | | | |

| | | | |
|---|---|---|---|
| **Tartaric Acid and its derivatives** | 2008148 | **Derivatives Reported with Piezoelectric Responses[71]** | **Max Predicted Strain: -21.47** |
| **L-Threonine[10]** | 5000016 | **0.1 ($d_{33}$)** | **0.14 ($d_{33}$)** |
| **N-methylurea** | 7209247 | **Previously has been gone under related piezoelectric study[72]** | |
| **Methylpiperazine** | 2105466 | **Has been reported as in chiral hybrid metal halides R/SMPCdCl4 (R/SMP = R-/S-2-methylpiperazine)[73]** | **-10.58 ($d_{14}$)** |
| **5-aminotetrazolium dinitramide** | 7000350 | **Derivatives reported as energetic materials[74]** | |
| **p-nitrobenzaldehyde** | 2005110 | **Reported for Electrical, Optical and Mechanical properties[75]** | |

**Table S6.** Summary of amino acid single crystal systems used for the comparison of calculated and experimental piezoelectric constants. Experimental and simulation data as reported in Table 1 of the ref.[76]

| Amino Acid Crystal Name | Piezoelectric Strain Constant ($d_{ij}$) | CrystalDFT Predictions (pC/N) | Experimental/Simulation Data (Literature) (pC/N) |
|---|---|---|---|
| L-Alanine | $d_{36}$ | 5.7 | 6.3 |
| L-Arginine | $d_{16}$ | 6.76 | 6.5 |
| L-Asparagine | $d_{36}$ | 11.02 | 13 |
| L-Aspartate | $d_{34}$ | 13.2 | 13 |
| L-Cysteine | $d_{22}$ | 10.19 | 11.4 |
| γ-Glycine[7] | $d_{33}$ | 10.72 | 11.33 |
| L-histidine | $d_{16}$ | 18.49 | 18 |
| L-Leucine | $d_{34}$ | 13.33 | 8.6 |
| L-Methionine | $d_{34}$ | 17.18 | 15 |
| L-Threonine | $d_{36}$ | 13.46 | 4.9 |
| L-Threonine | $d_{33}$ | 0.1 | 0.14 |
| DL-Alanine[77] | $d_{33}$ | 10.3 | 10.8 |
| DL-Tyrosine[78] | $d_{33}$ | 8.4 | 8.64 |